\documentclass[aps,prb,reprint,superscriptaddress]{revtex4-2}

\usepackage{amsmath, amssymb, amsthm, amsfonts} 
\usepackage{physics}
\usepackage{mathtools}
\usepackage{cases}

\usepackage{graphicx}
\usepackage{multirow}
\usepackage[dvipsnames]{xcolor} %
\usepackage{booktabs}
\usepackage{siunitx}

\usepackage{algorithm}
\usepackage{algpseudocode}

\usepackage[utf8]{inputenc}
\usepackage[T1]{fontenc}
\usepackage{xspace}
\usepackage{comment}
\usepackage{scalerel}
\usepackage{setspace}
\usepackage{booktabs, siunitx, xcolor, colortbl, rotating}
\definecolor{highlight}{RGB}{255, 255, 180}
\usepackage{hyperref}
\hypersetup{colorlinks=true, linkcolor=blue, citecolor=blue, urlcolor=blue}
\definecolor{americanrose}{rgb}{1.0, 0.01, 0.24}

\begin{document}

\preprint{APS/123-QED}

\title{Mitigating Noise-Induced Barren Plateaus Using a Non-Unitary 
Ansatz: Application to Molecular Electronic Transport}
\author{Sasanka Dowarah}
\email{sasanka.dowarah@utdallas.edu}
\affiliation{Department of Physics, The University of Texas at Dallas, Richardson, Texas 75080, USA}

\author{Abeda Sultana Shamma}
\email{Abeda.Shamma@UTDallas.edu}
\affiliation{Department of Physics, The University of Texas at Dallas, Richardson, Texas 75080, USA}
\affiliation{Department of Chemistry and Biochemistry, The University of Texas at Dallas, Richardson, Texas 75080, USA}

\author{Yazdan Maghsoud}
\email{ymaghsoud3@gatech.edu}
\thanks{Current address: School of Chemistry and Biochemistry, Georgia Institute of Technology, Atlanta, Georgia 30332, USA}
\affiliation{Department of Chemistry and Biochemistry, The University of Texas at Dallas, Richardson, Texas 75080, USA}

\author{G. Andrés Cisneros}
\email{andres@utdallas.edu}
\affiliation{Department of Physics, The University of Texas at Dallas, Richardson, Texas 75080, USA}
\affiliation{Department of Chemistry and Biochemistry, The University of Texas at Dallas, Richardson, Texas 75080, USA}

\author{Michael Kolodrubetz}
\email{mkolodru@utdallas.edu}
\affiliation{Department of Physics, The University of Texas at Dallas, Richardson, Texas 75080, USA}

\date{\today}

\begin{abstract}
Variational quantum algorithms (VQAs) offer a promising route toward simulating many-body quantum systems on noisy intermediate‑scale quantum (NISQ) hardware. However, their scalability is severely limited by noise‑induced barren plateaus (NIBPs), where hardware noise causes the gradients of the cost function to vanish exponentially with circuit depth, rendering optimization impossible. In this work, we demonstrate that introducing nonunitary elements into the variational ansatz can mitigate NIBPs in open-quantum systems. Using an analytically tractable infinite-range dissipative Ising model, we show that a nonunitary ansatz restores finite gradients in the presence of depolarizing noise, enabling convergence to the correct symmetry‑broken steady state. We also develop a Floquet‑type variational ansatz in which each layer repeats the same parameters, reducing the deep variational circuit to an effective quantum channel whose fixed points can be analyzed directly. We then extend these ideas to a realistic quantum‑chemistry system by simulating electron transport through Oligophenylethynylene-sulfurmethyl (OPE-SMe) using Hamiltonians and jump operators of the model derived from first‑principles polarizable QM/MM calculations. Our results show that nonunitary variational ans\"atze provide a scalable and physically grounded route for simulating open‑system steady states on NISQ hardware, offering a pathway to overcoming one of the limitations of current quantum hardware.
\end{abstract}
\maketitle

{\allowdisplaybreaks}
 \parskip 0 pt

\section{Introduction}
\begin{figure*}
    \centering
    \includegraphics[width=0.9\linewidth]{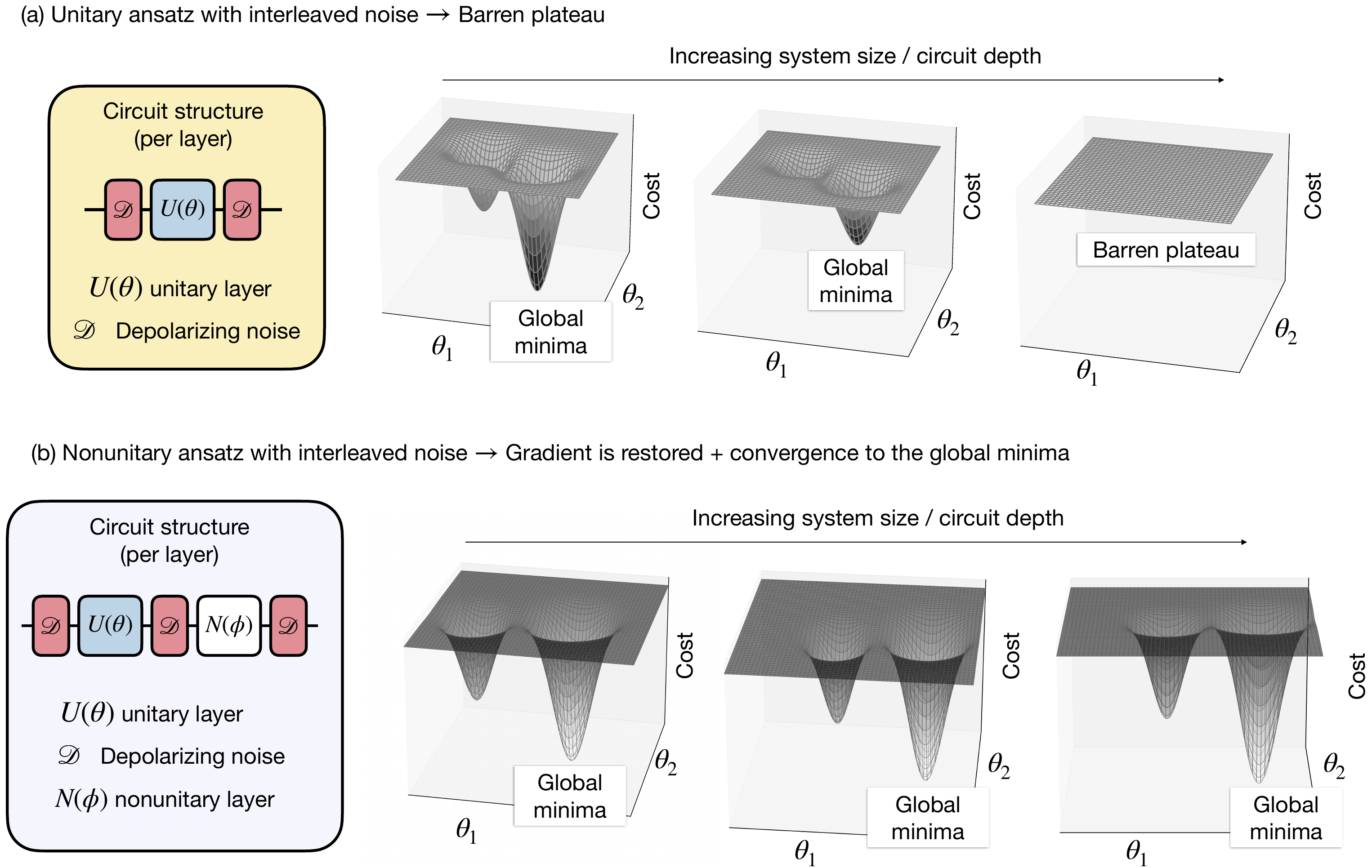}
    \caption{(a) Standard unitary variational ansatz with depolarizing noise interleaved between layers. As the system size or circuit depth increases, the cost‑function landscape flattens due to noise‑induced concentration of measure, causing gradients to vanish exponentially and leading to a barren plateau where optimization becomes impossible. (b) Nonunitary variational ansatz with interleaved noise, where each layer includes a controlled nonunitary channel $\mathcal{N}(\phi)$ in addition to unitary operations. The inclusion of nonunitary dynamics reshapes the effective optimization landscape, preventing gradient suppression and enabling convergence toward the global minimum even at large depth or system size. Note that the nonunitary ansatz does not merely restore gradients on the same optimization landscape; rather, the inclusion of nonunitary dynamics reshapes the cost landscape itself.}
    \label{fig:placeholder}
\end{figure*}
Variational quantum algorithms (VQAs) are among the most promising candidates for achieving quantum advantage on noisy intermediate-scale quantum (NISQ) devices \cite{Peruzzo_2014, McClean_2016, Cerezo_2021, Tilly_2022}. Their appeal comes from: (i) shallow, parameterized quantum circuits that can be implemented on near-term hardware, and (ii) a hybrid quantum--classical optimization loop that delegates the costly state preparation and measurement to the quantum processor while using classical routines to update parameters. VQAs have been successfully implemented in various settings from quantum chemistry and many-body physics, often with hardware-efficient and physics-informed ans\"atze tailored to specific platforms and problem structures \cite{TILLY20221, Kandala2017, PhysRevX.6.031007, Reiner_2019, Kokail2019}.

Despite this progress, the scalability of VQAs is limited by various factors, including ansatz expressibility and measurement overheads. Among these, one of the central obstructions is the barren plateau phenomenon, in which gradients of the cost function become exponentially suppressed, and optimization becomes impossible \cite{McClean2018, Cerezo2021}. While several mechanisms can contribute towards barren plateaus, including random parameter initialization, global cost functions, and overly expressive circuit families, the most severe limitation in the NISQ era is often hardware noise itself. It has been shown that in the presence of noise in the quantum circuits, the gradients vanish exponentially with the circuit depth, a phenomenon that was called noise-induced barren plateaus (NIBPs), and it is independent of initialization strategy or cost-function locality \cite{Wang2021}. A variety of strategies have been proposed to mitigate barren plateaus in noiseless or weak-noise settings, including layerwise training and problem-informed parameter initialization \cite{Grant2019initialization, Skolik2021}, the use of local cost functions \cite{Cerezo2021}, and structured ans\"atze designed to avoid effective randomness \cite{Huggins_2019, Larocca2025}. Error mitigation approaches can also improve trainability in specific regimes \cite{Wang_2024}. Another solution that has been proposed is to use nonuitary variational ansatz rather than unitary ones. While most of the variational quantum algorithms have been formulated in terms of purely unitary circuits, there has been growing interest in dissipative quantum computing and its potential to overcome trainability limitations \cite{Verstraete2009}. Recently, a number of works have investigated how non-unitary or non-unital channels can mitigate noise-induced barren plateaus \cite{zapusek2025scalingquantumalgorithmsdissipation, Sannia2024, Mele2026}. Despite this rapid development, their application to realistic, large-scale physical problems remains limited.
In this work, we fill this gap. We first show that introducing learnable nonunitary operations into the variational ansatz provides a physically motivated route to mitigating NIBPs, especially in simulating open-quantum systems. We consider variational circuits that interleave (i) coherent unitary generators with (ii) tunable dissipative channels that are constrained to be completely positive and trace preserving (CPTP). These nonunitary elements are part of the ansatz and are optimized alongside the unitary parameters, while the hardware noise is modeled separately as a fixed depolarizing channel inserted between circuit elements. We find that the resulting nonunitary ansatz can preserve parameter sensitivity and restore finite gradients in regimes where purely unitary ans\"atze exhibit NIBPs.

To make these statements precise and to retain analytical control, we first study an infinite-range transverse-field Ising model with dissipation. In the mean-field limit, the open-system dynamics reduces to a single-qubit Lindblad problem with a symmetry-broken steady state and a trivial steady state. This model serves as a benchmark: it exhibits a clear ferromagnetic--paramagnetic transition and allows direct comparison between (i) the exact mean-field steady state and (ii) the state reached by variational optimization in the presence of depolarizing noise. We demonstrate that a unitary ansatz suffers rapid gradient suppression with depth under depolarizing noise, while the nonunitary ansatz maintains trainability and converges to the correct symmetry-broken steady state. We further show that these conclusions hold for multiple choices of cost functions---a cost function that maximizes the energy of the system, and a cost function that gives the steady state of the system on minimization.

We also introduce a \emph{Floquet-type} construction in which all circuit layers share the same parameters. In this limit, the deep variational circuit reduces to an effective quantum channel whose fixed points can be analyzed directly. This viewpoint reveals how repeated unital noise drives parameter dependence to vanish in purely unitary circuits, and how tunable nonunitary components modify the channel fixed-point structure to preserve parameter sensitivity and prevent gradient collapse. We also show that the nonunitary ansatz converges to the global optima of the potential landscape, which corresponds to the steady state of the open-quantum system.

Finally, to demonstrate that the approach extends beyond analytically tractable toy models, we apply the same framework to steady‑state electron transport through an Oligophenylethynylene-sulfurmethyl (OPE-SMe) molecular junction. We model transport using a Lindblad description derived from first‑principles polarizable QM/MM calculations, with electron injection and extraction processes represented by fermionic jump operators acting on the molecular orbitals. We implement the resulting nonunitary dynamics using ancilla‑assisted quantum circuits and Trotterized time evolution, and we show that the variational nonunitary ansatz reliably converges to the nonequilibrium steady state even in the presence of noise. By comparing the resulting steady‑state charge currents across different solvent environments, we find that the variational approach reproduces the expected qualitative ordering of currents, demonstrating that the method captures the essential physics of solvent‑induced gating while remaining compatible with NISQ-era constraints.

The remainder of this paper is organized as follows. In Sec. II, we introduce the variational framework for open quantum systems, define the cost functions used to target nonequilibrium steady states, and review the origin of noise‑induced barren plateaus under unital noise. In Sec. III, we present results for the dissipative infinite‑range Ising model, including the mean‑field steady states, the nonunitary variational ansatz, numerical evidence for the mitigation of noise‑induced barren plateaus, and the Floquet‑channel mechanism underlying the observed trainability. In Sec. IV, we apply the method to steady‑state electron transport through OPE-SMe and discuss implementation details relevant for NISQ‑era hardware. Finally, we summarize our findings and provide an outlook in Sec. V.
\section{Variational quantum algorithms}
\subsection{Overview}
\begin{figure}
    \centering
    \includegraphics[width=\linewidth]{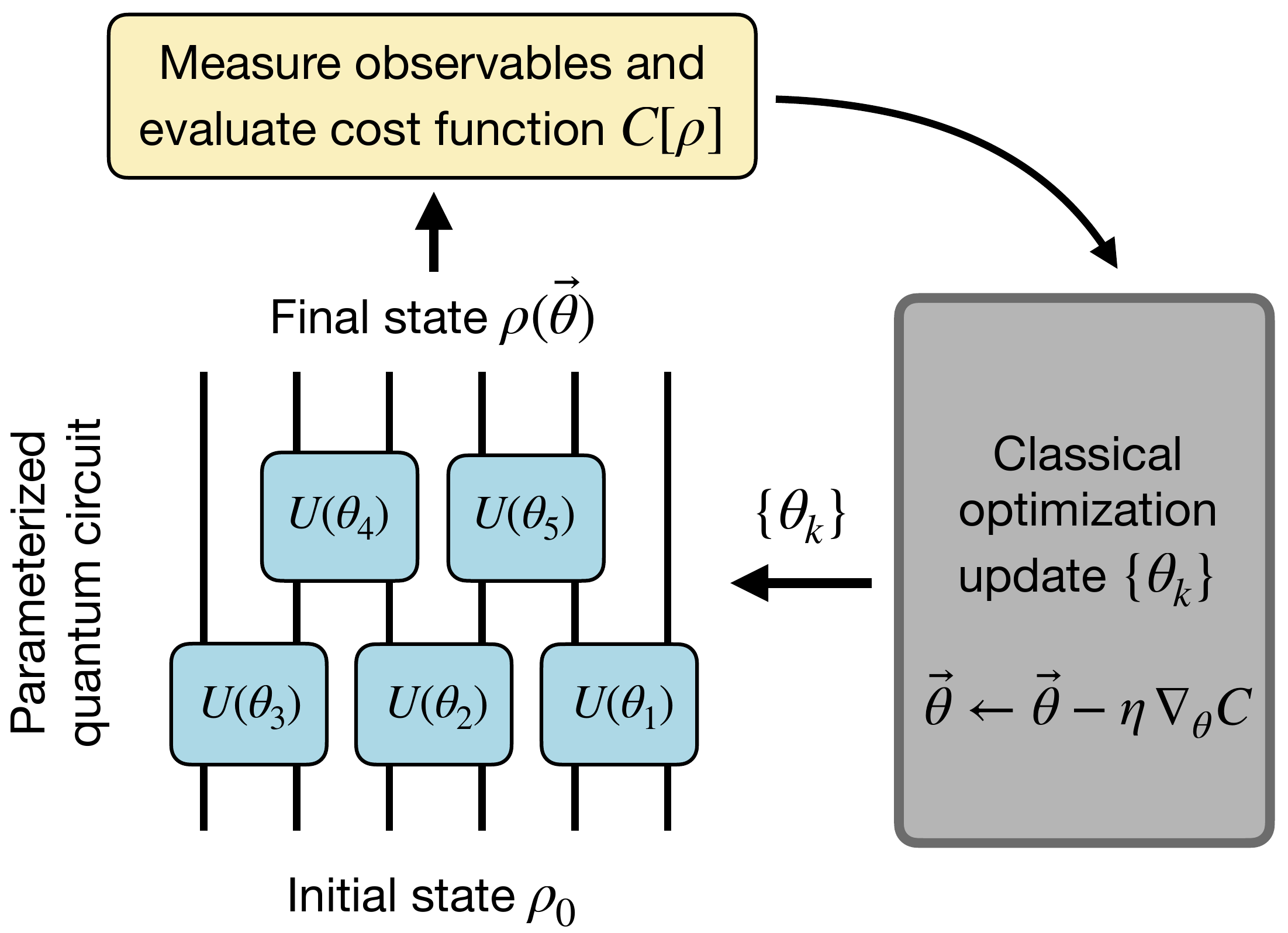}
    \label{fig:VQA-quantum-classical-loop-diagram}
    \caption{Schematic of a variational quantum algorithm. An initial quantum state $\rho_{0}$ is evolved by a parameterized quantum circuit composed of gates $U(\theta_{k})$, producing the trial state $\rho(\vec{\theta})$. Measurements of suitable observables are used to evaluate the cost function $C[\rho(\vec{\theta})]$, which is then supplied to a classical optimizer. The optimizer updates the variational parameters $\{\theta_{k}\}$, for example according to gradient descent rule $\vec{\theta}\leftarrow \vec{\theta}-\eta\nabla_{\vec{\theta}}C$, and the hybrid quantum--classical loop is repeated until convergence.}
\end{figure}
Variational quantum algorithms (VQA) are a broad class of hybrid quantum-classical algorithms designed to leverage near-term (noisy) quantum devices. A VQA optimizes a parameterized quantum circuit to minimize (or maximize) a cost function that can be evaluated efficiently on a quantum processor. The central task is to optimize a cost function of the form
\begin{eqnarray}
    C(\vec{\theta}) &=& \langle \psi(\vec{\theta})| \hat{O} | \psi (\vec{\theta})\rangle, \label{eq:Cost-function-definition-in-VQA}
\end{eqnarray}
where the variational state is prepared by $|\psi(\vec{\theta})\rangle = U(\vec{\theta}) |0\rangle^{\otimes L}$, $\vec{\theta} = (\theta_{1}, \theta_{2}, \cdots, \theta_{p})$ denotes a vectors with $p$ tunable parameters, and $\hat{O}$ is a Hermitian operator that encodes the goal of the problem, \emph{e.g.} if the goal is to find the ground state of a Hamiltonian then the $\hat{O}$ is the Hamiltonian such that minimizing Eq.~\eqref{eq:Cost-function-definition-in-VQA} minimizes the energy.

For the Hermitian operator $\hat{O}$, it can be written in terms of local or few-body observables:
\begin{eqnarray}
    \hat{O} &=& \sum_{j} c_{j} P_{j},
\end{eqnarray}
where each $P_{j}$ is a Pauli operator acting on a subset of qubits. The cost function can then be written as
\begin{eqnarray}
    C(\vec{\theta}) &=& \sum_{j} c_{j} \langle \psi(\vec{\theta})|P_{j}|\psi(\vec{\theta})\rangle. \label{eq:Cost-function-using-Pauli-observables}
\end{eqnarray}
Each of these terms $\langle \psi(\vec{\theta})|P_{j}|\psi(\vec{\theta})\rangle$ can be measured on a quantum processor efficiently. These measurement results are then processed in a classical computer to evaluate the cost function $C(\vec{\theta})$ and to update the parameters for the next iteration. This update can be done using gradient descent
\begin{eqnarray}
    \theta^{(n + 1)}_{k} &=& \theta^{(n)}_{k} - \eta \frac{\partial C}{\partial \theta_{k}}, \label{eq:Gradient-descent-rule-for-parameter-update}
\end{eqnarray}
where $\eta$ is the learning rate.

The gradient of the cost function can be evaluated using the parameter shift rule, which for gates of the form $U(\theta_{k}) = e^{- i \theta_{k} G_{k} / 2}$, with $G_{k}$ a Pauli generator, becomes \cite{PhysRevA.98.032309}
\begin{eqnarray}
    \frac{\partial C}{\partial \theta_{k}} &=& \frac{1}{2}\Bigg[ C\bigg(\vec{\theta} + \frac{\pi}{2}\vec{e}_{k}\bigg) - C\bigg(\vec{\theta} - \frac{\pi}{2} \vec{e}_{k}\bigg) \Bigg], \label{eq:Parameter-shift-rule-for-Pauli-gates}
\end{eqnarray}
which allows calculation of the gradients with two additional evaluations of the cost function on the quantum device. This continues until a desired accuracy is achieved in terms of the cost function of the problem.

In practice, the number of layers required for a variational circuit generally increases with the system size---at least linearly or in some cases polynomially. Thus, as the size of the Hilbert space grows, it makes the variational circuit deeper. This increasing depth makes the cost landscape increasingly sensitive to parameter change and noise. This leads to the phenomenon of noise-induced barren plateaus, which we discuss in the next section.

While most quantum algorithms have been developed for closed unitary dynamics, many problems in chemistry, condensed matter physics, and quantum technologies involve inherently open-system behavior. In Markovian systems, a central task is to prepare the steady state of the Liouvillian operator $\mathcal{L}\rho_{\mathrm{ss}} = 0$. Several methods have been developed towards achieving this \cite{Delgado-Granados2025, watad2023variationalquantumalgorithmssimulation, Huang_2025, Luo2024, Santos2025lowrankvariational, b8tq-169m, Liu_2021}. But despite these advances, the feasibility of preparing steady state variationally on realistic NISQ hardware is still not well understood, and the role of NIBPs in this setting remains essentially unexplored.
\subsection{Noise-induced barren plateau and local minima}
For a unitary ansatz, prior work shows that in the case of local Pauli noise channels acting before and after every variational layer, whose action on a local Pauli operator $\sigma \in \{ \sigma^{x}, \sigma^{y}, \sigma^{z}\}$ is given by $\mathcal{N}_{j}(\sigma) = q_{\sigma} \sigma$, where $-1 < q_{X}, q_{Y}, q_{Z} < 1$, the magnitude of the gradient of typical VQA cost functions vanishes exponentially with circuit depth $L$
\begin{eqnarray}
    \bigg|\frac{\partial C}{\partial \theta_{lm}}\bigg| \sim e^{ - c L}, \quad c > 0,
\end{eqnarray}
where $\theta_{lm}$ denotes the $m-$th parameter from the $l-$th layer \cite{Wang2021}. Since depth often grows at least linearly with system size, this leads to a flat, featureless optimization landscape and untrainability. Unlike noise-free barren plateaus, where vanishing of the gradient depends on random initialization or non-locality of the cost-function, the noise-induced barren plateau (NIBP) arises deterministically from the dissipative dynamics of the noisy quantum channels. It is therefore independent of the initialization strategy \cite{McClean2018} or cost-function locality.

 Beyond barren plateaus, the optimization landscape of a variational quantum algorithm may contain numerous local minima that degrade the quality of the solution \cite{PhysRevLett.127.120502, Larocca2025}. In high-dimensional parameter spaces, the optimizer may become trapped in such minima, producing solutions that approximate suboptimal excited states rather than the true ground state or the optimal cost value. Avoiding these local minima is therefore crucial to ensure the robustness, accuracy, and expressibility of the ansatz. The problem becomes harder in the presence of noise, which can significantly alter the cost landscape by either smoothing or distorting it. Together with mitigating the NIBPs, avoiding such local minima is essential for maintaining reliable trainability and scalable performance of the variational quantum algorithms.

 In the next section, we show that introducing a nonunitary ansatz can mitigate NIBPs. To demonstrate these effects and to maintain analytical control, we consider an all-to-all interacting dissipative Ising model that is solvable in the mean-field limit.

\section{Results}
\subsection{Benchmark: dissipative infinite-range Ising model}
\begin{figure}
    \centering
    \includegraphics[width=0.99\linewidth]{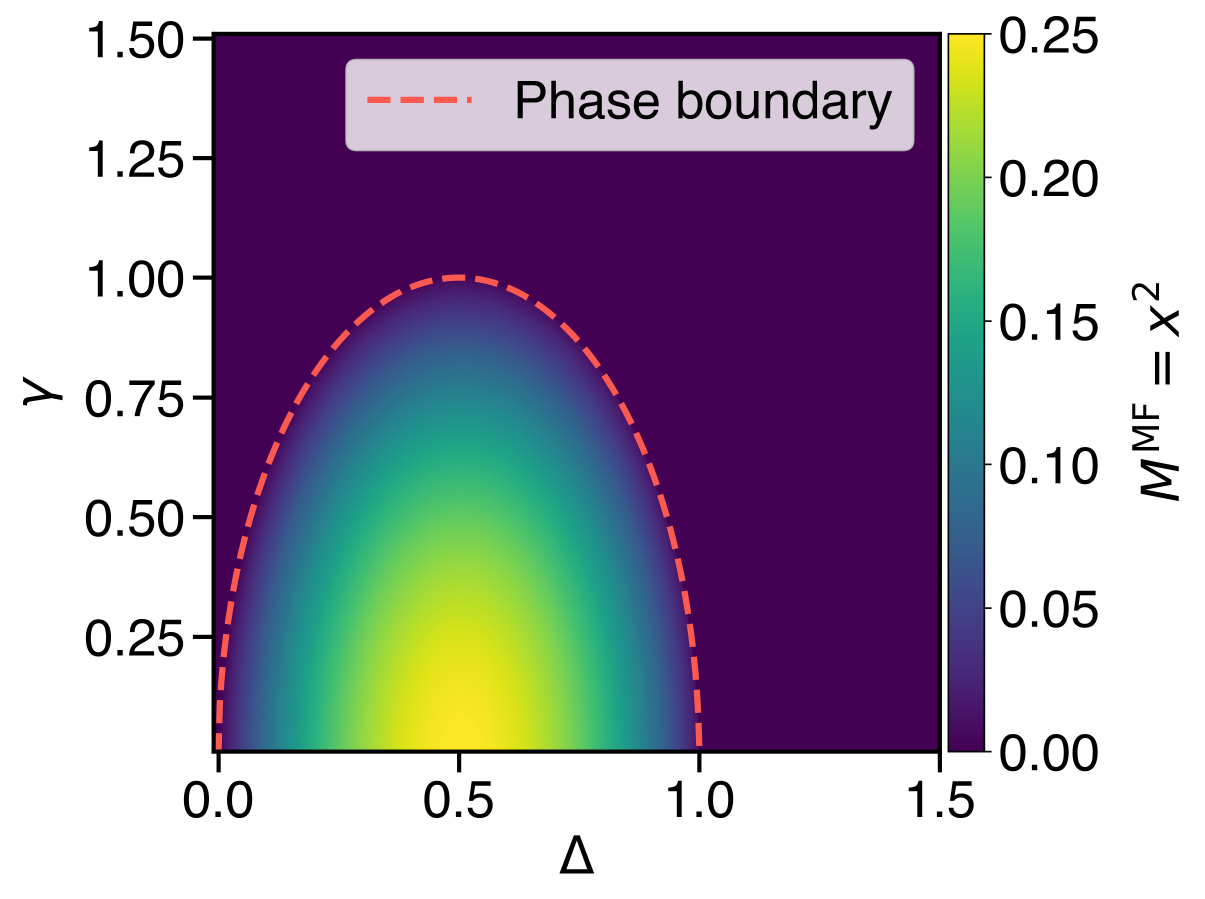}
    \caption{Steady state phase diagram with the mean-field order parameter as a function of dissipation rate $\gamma$ and transverse field strength $\Delta$. The color represents the value of the mean-field order parameter $M_{\mathrm{F}} = x^{2}$. The red dashed line shows the analytically predicted phase boundary using Eq.~\eqref{eq:Phase_boundary_between_FM_PM}.}
    \label{fig:Phase_diagram}
\end{figure}

To benchmark trainability in a setting where the steady state is known analytically, we consider an infinite-range transverse-field Ising model with Markovian dissipation on each spin. We write the Hamiltonian with $N$ spins as follows
\begin{eqnarray}
    H &=& \sum_{i < j} J_{i j}\sigma^{x}_{i} \sigma^{x}_{j} - \Delta \sum_{i} \sigma^{z}_{i}, \label{eq:TFIM_model}
\end{eqnarray}
where each spin is subjected to two dissipative processes represented by two jump operators: relaxation $L_{\mathrm{rel}}= \sqrt{\gamma_{e}} |1\rangle \langle 0| = \sqrt{\gamma_{e}} \sigma^{-}$, and dephasing $L_{\mathrm{dep}} = \sqrt{\gamma_{d}} |0\rangle\langle 0| = \sqrt{\gamma_{d}} (I + \sigma^{z}) / 2$. We will set all coupling constants to be equal $J_{ij} = J$. This model has the following order parameter to detect the ferromagnetic-paramagnetic (FM-PM phase) transition \cite{haack2023probingnonequilibriumdissipativephase}:
\begin{eqnarray}
    M_{F} &=& \sum_{i, j} \frac{\langle \sigma^{x}_{i} \sigma^{x}_{j} \rangle}{N^{2}}. \label{eq:Order-parameter-for-FM-PM-phase-transition}
\end{eqnarray}
\textit{Mean-field solution} ---
We now use mean-field theory to simplify the Hamiltonian and map out its phase diagram. Assuming that the density matrix has permutation symmetry among the spins, the interacting part of the Hamiltonian can be simplified as
\begin{eqnarray}
   \sum_{i < j} J \sigma^{x}_{i} \sigma^{x}_{j} \approx J N \sum_{i} \sigma^{x}_{i} \langle \sigma^{x} \rangle.
\end{eqnarray}
From now on, we will set $JN = 1$. With this, the mean-field Hamiltonian on each spin is given by
\begin{eqnarray}
    H_{\mathrm{MF}} &=& \sigma^{x} \langle \sigma^{x} \rangle - \Delta \sigma^{z} \label{eq:MF_Hamiltonian}
\end{eqnarray}
The ferromagnetic order parameter in the mean-field limit becomes
\begin{eqnarray}
    M_{\mathrm{F}} \approx \langle \sigma^{x}\rangle^{2} \equiv x^{2}. \label{eq:Order_parameter_mean_field_limit}
\end{eqnarray}
In order to calculate the steady state phase diagram for this mean-field model, let us assume that the density matrix of a single qubit has the form
\begin{eqnarray}
    \rho &=& \frac{1}{2} (I + x \sigma^{x} + y \sigma^{y}
    + z \sigma^{z}), \label{eq:Density_matrix}
\end{eqnarray}
where $x = \mathrm{Tr}[\sigma^{x}\rho], y = \mathrm{Tr}[\sigma^{y}\rho]$ and $z = \mathrm{Tr}[\sigma^{z}\rho]$. For self-consistency, we must have $\langle \sigma^{x} \rangle = x$ in Eq.~\eqref{eq:MF_Hamiltonian} and ~\eqref{eq:Order_parameter_mean_field_limit}. In this mean-field approximation, every qubit in the system has the same density matrix given by Eq.~\eqref{eq:Density_matrix}. We now solve the evolution of the Bloch components $(x, y, z)$ of the density matrix to obtain the Optical Bloch equations $($OBE$)$
\begin{subequations}
\begin{eqnarray}
    \frac{d x}{d t} &=& 2 \Delta y - \frac{\gamma_{e}}{2} x - \frac{\gamma_{d}}{2} x, \\
    \frac{d y}{d t} &=& - 2 z x - 2 x \Delta  - \frac{\gamma_{e}}{2} y - \frac{\gamma_{d}}{2} y, \\
    \frac{d z}{d t} &=& 2 y x - \gamma_{e} ( 1 + z). 
\end{eqnarray}\label{eq:Optical_Bloch_Equations}
\end{subequations}
In the steady state $\frac{d x}{d t} = \frac{d y}{d t} = \frac{d z}{d t} = 0$. Solving Eq.~\eqref{eq:Optical_Bloch_Equations}, we obtain two steady state solutions. The first, non-trivial solution is
\begin{eqnarray}
x_{\mathrm{ss}} &=& \pm \frac{4 \Delta}{\gamma_{e} + \gamma_{d}} \sqrt{P}, \;
    y_{\mathrm{ss}} = \pm \sqrt{P}, \;
    z_{\mathrm{ss}} =  Q, \label{eq:Steady-state-components-in-FM-phase}
\end{eqnarray}
where
\begin{subequations}
\begin{eqnarray}
    P &=& \frac{ \gamma_{e} (\gamma_{e} + \gamma_{d} )}{ 8 \Delta} (1 + Q), \\
    Q &=& \frac{- 16 \Delta^{2} - (  \gamma_{e} + \gamma_{d})^{2} }{16 \Delta}.
\end{eqnarray}
\end{subequations}
The second, trivial solution is
\begin{eqnarray}
    x_{\mathrm{ss}} = 0, \; y_{\mathrm{ss}} = 0, \; z_{\mathrm{ss}} = - 1. \label{eq:PM_phase_solution_steady_state}
\end{eqnarray}
The first solution in Eq.~\eqref{eq:Steady-state-components-in-FM-phase} corresponds to the FM phase, which has a nonzero $x$ parameter, and the second (Eq.~\eqref{eq:PM_phase_solution_steady_state}) corresponds to the PM phase, where the steady state $x$ parameter vanishes. From the requirement that $x_{\mathrm\
ss}$ and $y_{\mathrm{ss}}$ must be real in Eq.~\eqref{eq:Steady-state-components-in-FM-phase}, we get the phase boundary between FM and PM phase within the mean-field limit:
\begin{eqnarray}
   \gamma_{e} + \gamma_{d} &=& 4 \sqrt{\Delta (1 - \Delta )}. \label{eq:Phase_boundary_between_FM_PM}
\end{eqnarray}
In Fig.~\ref{fig:Phase_diagram}, we plot the steady-state order parameter in Eq.~\eqref{eq:Order_parameter_mean_field_limit} as a function of $\gamma = \gamma_{e} = \gamma_{d}$ and $\Delta$, along with the phase boundary predicted by Eq.~\eqref{eq:Phase_boundary_between_FM_PM} which shows the validity of the mean-field approximation. The authors in~\cite{haack2023probingnonequilibriumdissipativephase} obtain the same phase diagram for the model using a cumulant method that allows calculation of more complex correlation functions.

This model provides a rigorous testbed for our work: a successful variational procedure must converge to a symmetry-broken steady state with $x_{\mathrm{ss}} \neq 0$ in the ferromagnetic regime while remaining robust to hardware noise. Having established the structure of the exact steady state in the mean-field limit, we now construct a variational ansatz for the steady state of this model.

\subsection{Variational ansatz for the Ising model with depolarizing noise}
We now describe the variational ansatz used throughout the work for the mean-field Hamiltonian with dissipation. We define the nonunitary ansatz with $L$ layers as
\begin{eqnarray}
    \mathcal{U}_{L} (\vec{\theta}) & = & \prod^{L}_{l= 1} \mathcal{U}_{l}(\theta^{(l)}_{x}, \theta^{(l)}_{\mathrm{dep}}, \theta^{(l)}_{z},
    \theta^{(l)}_{\mathrm{rel}},
    \theta^{(l)}_{y}),
\end{eqnarray}
where the $l$th layer is given by
\begin{eqnarray}
  \mathcal{U}_{l} (\vec{\theta}^{(l)}) & \equiv & 
    e^{\theta^{(l)}_{z} \mathcal{L}_{z}} \circ 
    e^{\theta^{(l)}_{\mathrm{rel}} \mathcal{L}_{\mathrm{rel}}} \circ
    e^{\theta^{(l)}_{y}\mathcal{L}_{y}} \circ
    e^{\theta^{(l)}_{\mathrm{dep}} \mathcal{L}_{\mathrm{dep}}} \circ 
    e^{\theta^{(l)}_{x}\mathcal{L}_{x}}. \nonumber
\end{eqnarray}
Here the symbol $\circ$ denotes composition of quantum channels, $(\mathcal{A} \circ \mathcal{B}) (\rho) = \mathcal{A}[\mathcal{B}(\rho)]$.
The unitary generators in the ansatz are given by
\begin{subequations}
\begin{eqnarray}
\mathcal{L}_{x} &=&-i x (I\otimes \sigma^{x}- \sigma^{x} \otimes I),\\
    \mathcal{L}_{z} &=&-i(I\otimes \sigma^{z} - \sigma^{z}\otimes I),\\
    \mathcal{L}_{y} &=& - i (I \otimes \sigma^{y} - \sigma^{y} \otimes I),
\end{eqnarray}
\end{subequations}
and the dissipators are given by
\begin{eqnarray}
    \mathcal{L}_{k} &=& L^{*}_{k} \otimes L_{k} - \frac{1}{2} I \otimes (L^{\dagger}_{k} L_{k}) - \frac{1}{2} (L^{\dagger}_{k} L_{k})^{T} \otimes I, \nonumber
\end{eqnarray}
where $L_{k}$ is  $L_{\mathrm{rel}} = \sigma^{-}$ or $ L_{\mathrm{dep}} = (I + \sigma^{z})/ 2$. This ansatz is chosen so that the exact Trotterized dynamics can be obtained in the limit of small $\theta$. In other words, our ansatz the dissipative generalization of the quantum adiabatic optimization algorithm (QAOA)~\cite{farhi2014quantumapproximateoptimizationalgorithm}. We impose $\theta_{\mathrm{rel}}, \theta_{\mathrm{dep}} \geq 0$ such that these nonunitary operators remain completely positive trace preserving $($CPTP$)$ maps. With $\theta_{\mathrm{rel}} = \theta_{\mathrm{dep}} = 0$, we obtain the unitary ansatz. 

\begin{figure*}[htp]
    \centering
    \includegraphics[width=\linewidth]{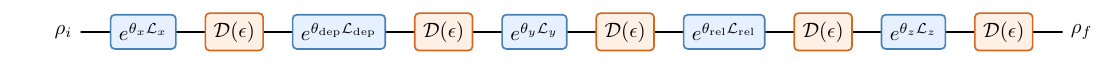}
    \caption{A single-layer variational quantum circuit for the mean-field model with each unitary and non-unitary gates sandwiched between two depolarizing noise channels.}
    \label{fig:Mean-field_model_ansatz_variaitonal_quantum_circuit}
\end{figure*}

\textit{Noise model} --- We consider a noise model where each unitary and non-unitary gate is sandwiched by a depolarizing channel $\mathcal{D}(\epsilon)$ of strength $\epsilon$---consistent with the noise model considered in \cite{Wang2021}. The depolarizing noise channel with strength $\epsilon$ is given by the following \cite{Nielsen_Chuang_2010}:
\begin{eqnarray}
    \mathcal{D}[\rho](\epsilon) &=& (1- \epsilon) \rho + \frac{\epsilon}{3} (\sigma_{x} \rho \sigma_{x} + \sigma_{y} \rho \sigma_{y}
    + \sigma_{z} \rho \sigma_{z}),\hspace{6 mm}
\end{eqnarray}
which can be interpreted as having probability $1- \epsilon$ of leaving the state unaffected, and a probability of $\epsilon/ 3$ of relaxing to the infinite temperature state $\rho_{\infty} = I/N$. With this noise model, the ansatz for the $l$-th layer becomes
\begin{widetext}
    \begin{eqnarray}
    \mathcal{U}_{l} (\vec{\theta}^{(l)}, \epsilon) &=& 
    \mathcal{D}(\epsilon) \circ
    e^{\theta^{(l)}_{z} \mathcal{L}_{z}} \circ
    \mathcal{D}(\epsilon)
    \circ e^{\theta^{(l)}_{\mathrm{rel}} \mathcal{L}_{\mathrm{rel}}} \circ \mathcal{D}(\epsilon) \circ
    e^{\theta^{(l)}_{y}\mathcal{L}_{y}}
    \circ \mathcal{D}(\epsilon) \circ
    e^{\theta^{(l)}_{\mathrm{dep}} \mathcal{L}_{\mathrm{dep}}}
    \circ \mathcal{D}(\epsilon) \circ
    e^{\theta^{(l)}_{x}\mathcal{L}_{x}} \label{eq:Variational_ansatz_with_noise_definition}
\end{eqnarray}
A schematic of this single-layer quantum circuit corresponding to Eq.~\eqref{eq:Variational_ansatz_with_noise_definition} is shown in Fig.~\ref{fig:Mean-field_model_ansatz_variaitonal_quantum_circuit}
\end{widetext}

\textit{Cost function} --
In an open quantum system, the true steady state $\rho_{\mathrm{ss}}$ is defined by the equation
\begin{eqnarray}
    \mathcal{L} \rho_{\mathrm{ss}} &=& 0,
\end{eqnarray}
where $\mathcal{L}$ is the Lindbladian superoperator of the system. Therefore, a variational algorithm that prepares the steady‑state must, in principle, minimize a quantity that vanishes exactly at the point $\rho = \rho_{\mathrm{ss}}$. For this purpose, we use the Frobenius‑norm cost function 
\begin{eqnarray}
    C_{\mathrm{F}}[\rho] = \mathrm{Tr}[(\mathcal{L}\rho)^{\dagger}(\mathcal{L}\rho)], \label{eq:Frobenius-norm-cost-function}
\end{eqnarray}
whose global minimum corresponds uniquely to the Lindblad steady state. This cost function provides a direct way to prepare the steady state but it is physically less transparent and more demanding computationally. 
Although $C_{\mathrm{F}}[\rho]$ is nonlinear in the density matrix and is therefore more demanding to evaluate than an ordinary expectation-value cost function, recent work has shown that Liouvillian-residual cost functions of this type can be estimated using classical-shadow techniques~\cite{zhou2023hybridalgorithmsimulatingnonequilibrium, Huang_2020}.

The Frobenius norm cost function is not in line with the one used in~\cite{Wang2021}, which require a form $C[\rho] = \mathrm{Tr}[O\rho]$ that is linear in $\rho$. To benchmark the analytically tractable mean-field Ising model, we use an additional and simple cost function that corresponds to the energy of the system
 \begin{eqnarray}
    C_{E}[\rho] &=& \mathrm{Tr}[H \rho] .\label{eq:Energy_cost_function_Ising_model}
\end{eqnarray}
For the mean-field model, this simplifies to
\begin{eqnarray}
    C_{E}[\rho] &=& \mathrm{Tr}[(x \sigma^{x} - \Delta \sigma^{z}) \rho] = x^{2} - \Delta z,
\end{eqnarray}
where $x, z$ are the components of the density matrix defined in Eq.~\eqref{eq:Density_matrix}. In order to obtain the ferromagnetic (FM) phase where $x_{\mathrm{ss}} \neq 0$, we maximize the energy cost function using gradient ascent.  
If instead, the cost function is minimized, the optimization converges to the trivial paramagnetic (PM) solution with $x_{\mathrm{ss}} = 0$. Although maximizing the energy is not the correct condition for the actual steady state of the problem, we show numerically that maximizing $C_{\mathrm{E}}$ selects the same symmetry‑broken solution as minimizing the Frobenius norm in Eq.~\eqref{eq:Frobenius-norm-cost-function}. This makes the energy cost function a valuable diagnostic tool: it provides an analytically interpretable cost landscape that allows us to probe trainability, expressibility, and the ability of the ansatz to locate the correct ferromagnetic phase without explicitly referencing the Liouvillian superoperator. In the next section, we numerically show that when the variational ansatz is nonunitary both the energy and the Frobenius norm cost functions remain trainable in the presence of depolarizing noise.

\subsection{Absence of noise-induced barren plateaus}
\begin{figure}
    \centering
    \includegraphics[width=\linewidth]{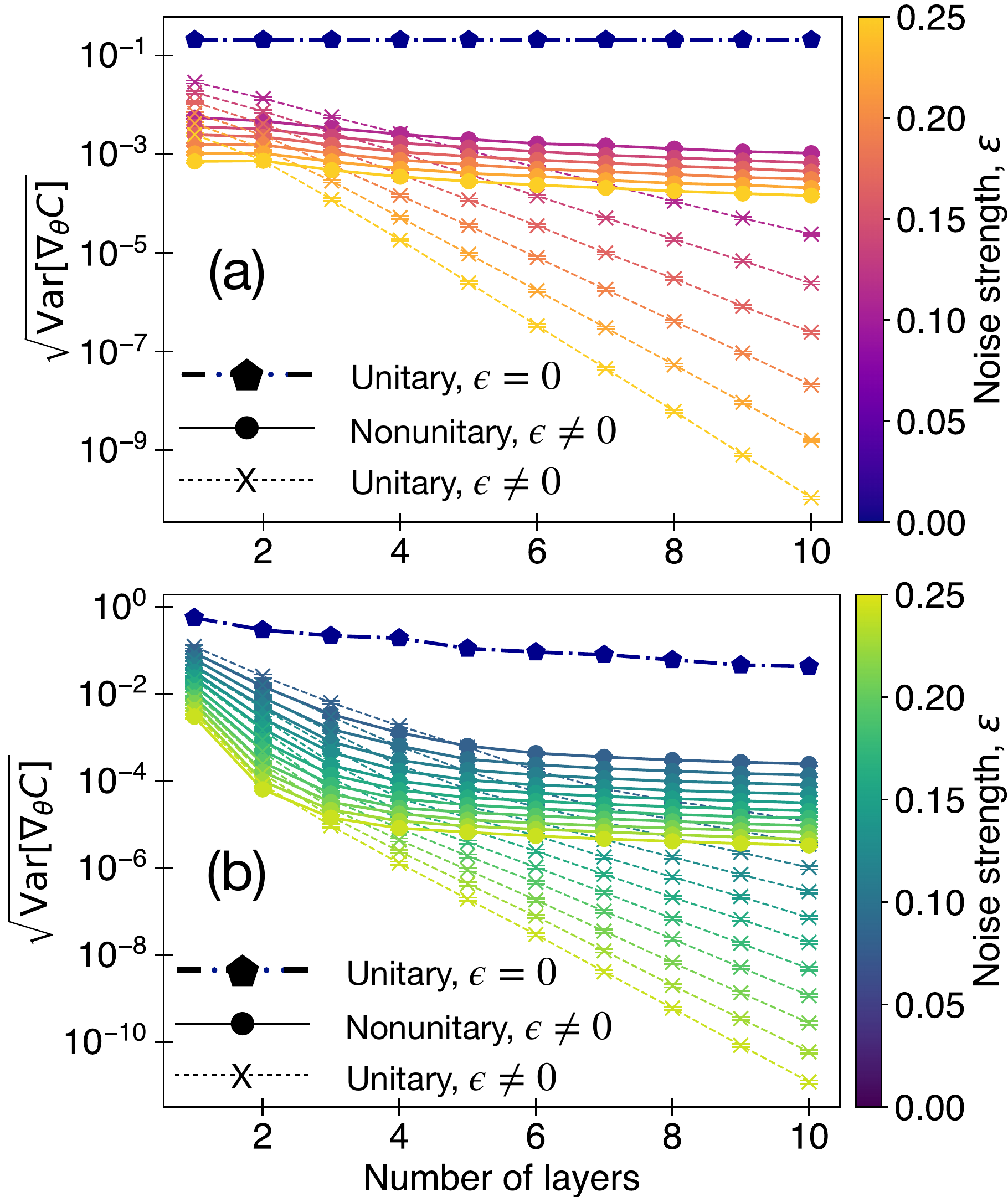}
    \caption{Plot of standard deviation of the gradients averaged over each layer and each parameters, as a function of number of layers for the unitary with no noise, unitary with noise and the nonunitary ansatz for different noise strengths, for the (a) energy cost function with $\Delta = 0.2$ and the (b) Frobenius norm cost function for $\Delta = 0.5, \gamma = 0.1$. Error bars indicate the standard error of the mean for $128$ random initial angles.}
    \label{fig:Gradient-vs-number-of-layers}
\end{figure}
We now show that the nonunitary ansatz can mitigate the barren plateaus induced by depolarizing noise, for both cost functions---energy and Frobenius norm. We initialize the system at $\rho_{i} = (I+\sigma^{y})/2$ for both the cost functions, and sample the initial unitary angles uniformly from $[-\pi, \pi]$, and the dissipative angles from $[0, 2\pi]$. We then calculate the gradients for one iteration and average the results over many initial angles. Our results are shown in Fig.~\ref{fig:Gradient-vs-number-of-layers}, where we plot the variance of the gradients of the cost function as a function of number of layers in the ansatz for both the unitary and the nonunitary cases. For the unitary ansatz, the gradient decreases exponentially with increasing circuit depth, indicating the onset of the NIBP phenomenon and the corresponding loss of trainability. In contrast, the nonunitary Floquet ansatz maintains a finite gradient even at large depths, demonstrating that the optimization landscape remains trainable in the presence of depolarizing noise. This behavior occurs for both the energy-based and Frobenius-norm cost functions, indicating that the mitigation mechanism is not tied to a particular choice of cost function but instead originates from the nonunitary structure of the ansatz itself.

 While the gradient analysis demonstrates that the nonunitary ansatz avoids noise‑induced barren plateaus, this alone is not sufficient to guarantee that the variational algorithm converges to the correct steady state. In principle, a finite gradient merely ensures that the parameters remain trainable; it does not preclude the optimizer from converging to spurious fixed points or suboptimal local extrema. Unlike many VQA settings, the present mean-field Ising model is effectively reduced to a single-qubit problem and therefore does not possess a natural notion of scaling with system size. In standard many-body variational circuits, increasing the system size generally requires increasing the circuit depth in order to maintain sufficient expressibility, making it possible to analyze how convergence and gradient behavior evolve with depth. In the present case, however, the effective mean-field dynamics are already encoded within a single-qubit description, so the gradient statistics by themselves do not completely determine whether the ansatz captures the correct physical steady state. To address this, in the next section, we employ a Floquet‑type ansatz in which a single parameterized layer is repeated indefinitely, reducing the problem to analyzing the fixed‑point structure of an effective quantum channel. Since the system is nonunitary, a Floquet ansatz has the potential to go to a single steady state. Therefore, this enables a simple example where we can analyze the deep circuit limit. We show that the resulting Bloch‑vector component $x$ (hence the order parameter in Eq.~\eqref{eq:Order_parameter_mean_field_limit}) converges to the ferromagnetic mean‑field value, demonstrating that the nonunitary ansatz not only remains trainable but also targets the symmetry-broken ferromagnetic phase.
\subsection{Floquet type ansatz}
\begin{figure}
    \centering
    \includegraphics[width=0.99\linewidth]{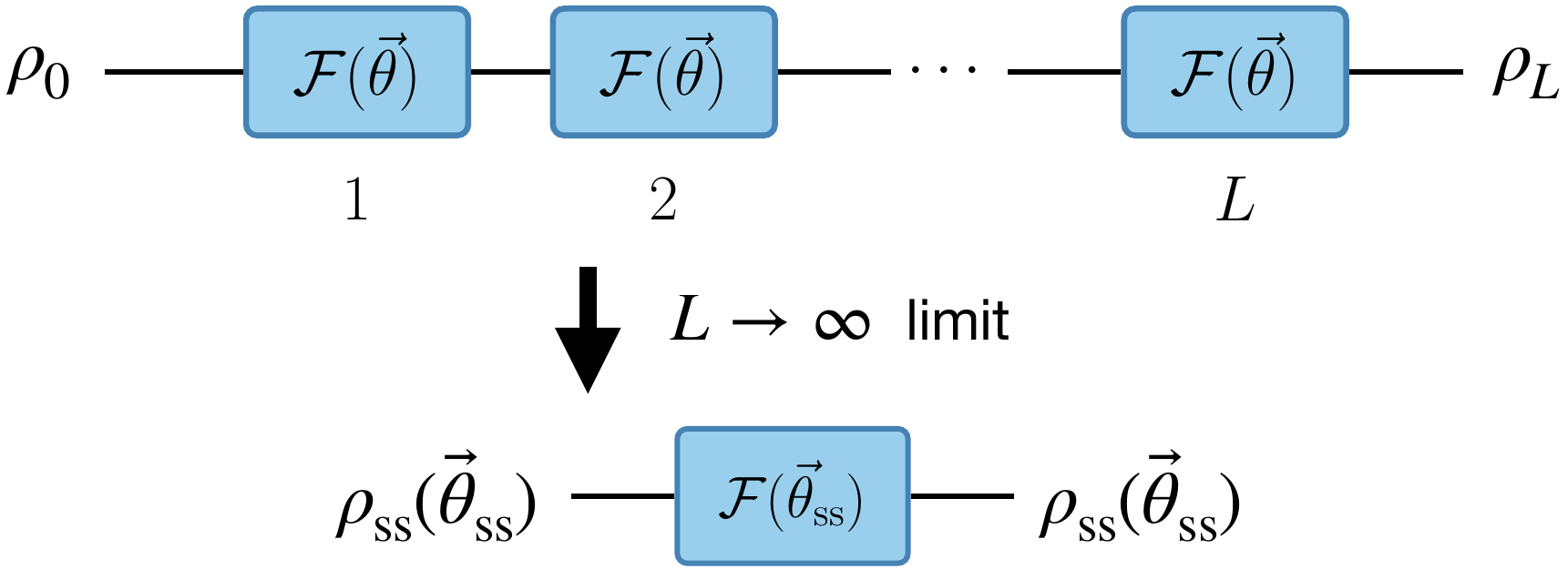}
    \caption{Floquet variational ansatz where each layer shares the same set of parameters. Repeating a single parameterized variational layer defines an effective Floquet superoperator that governs the long‑depth behavior of the circuit. In the infinite depth limit, the steady state of the system can be interpreted as a fixed point of the Floquet superoperator. In the presence of noise and purely unitary parameters, this channel possesses a unique trivial fixed point $\rho_{\mathrm{ss}}  = I/2$, leading to complete loss of parameter dependence. Introducing trainable nonunitary elements reshapes the fixed‑point structure of the channel, enabling nontrivial steady states to survive even in the presence of noise.}
    \label{fig:placeholder}
\end{figure}
\begin{figure*}
    \centering
    \includegraphics[width=\linewidth]{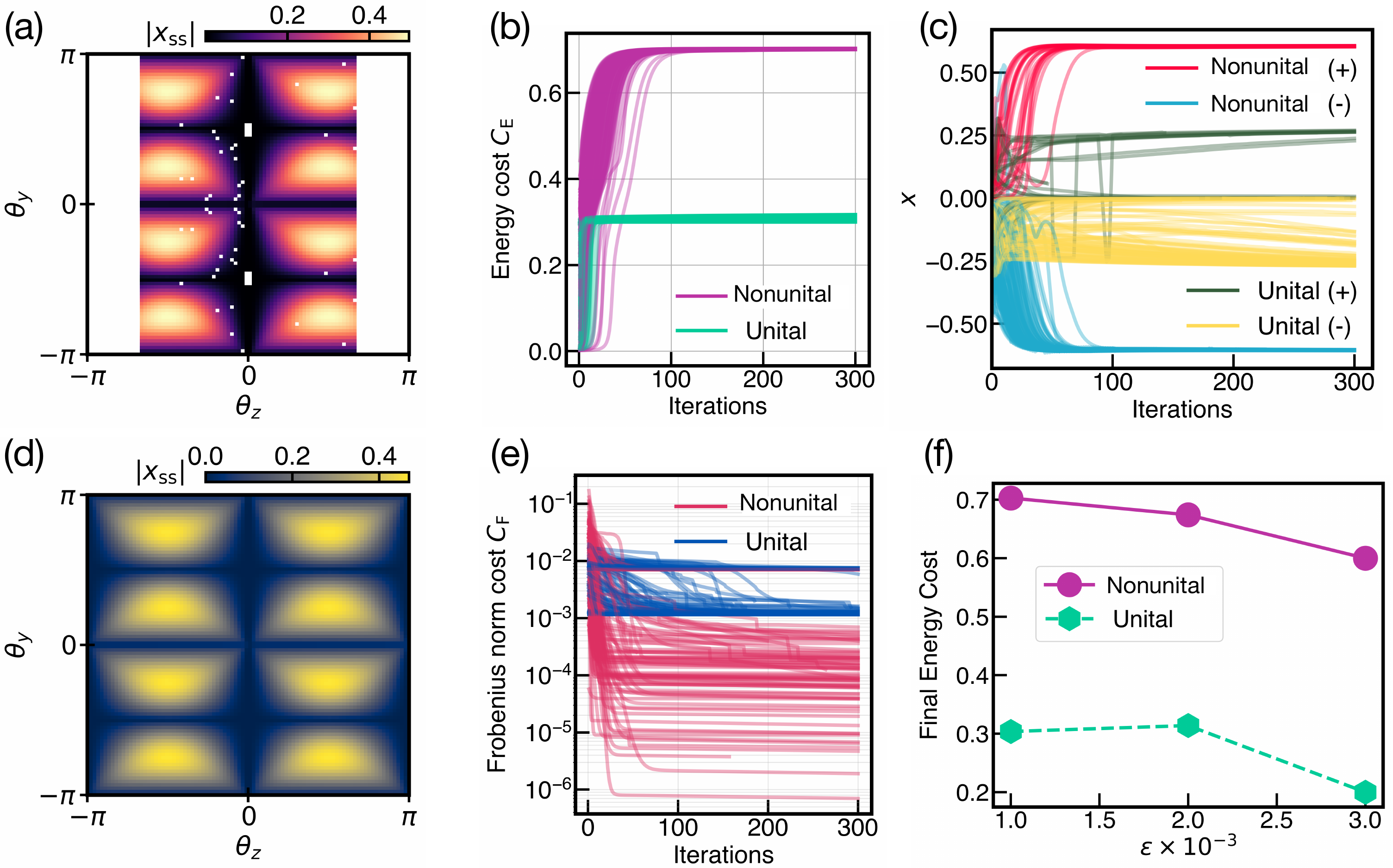}
    \caption{(a) Heatmap of the steady-state order-parameter magnitude $|x_{\mathrm{ss}}|$ as functions of the unitary angles $\theta_y$ and $\theta_z$ for the energy-cost function with $\Delta = 0.5$, $\theta_{\mathrm{dep}}=\theta_{\mathrm{rel}}=10^{-3}$, and $\epsilon=10^{-6}$. The plot shown for the value of $\theta_x$ for which $|x_{\mathrm{ss}}|$ is maximal. In the absence of dissipative channels ($\theta_{\mathrm{dep}}=\theta_{\mathrm{rel}}=0$) with $\epsilon \neq 0$, the dynamics relaxes to the trivial infinite-temperature steady state $\rho_{\mathrm{ss}}=I/2$, corresponding to $x_{\mathrm{ss}}=0$. The missing data points in the plots represents the angles where the ansatz failed to converge to a steady state.
    (b) Variational optimization of the energy-cost function as a function of iteration number for $\Delta = 0.5$ and $\epsilon = 10^{-3}$, comparing the nonunital and unital ansätze. In the unital ansatz, $\theta_{\mathrm{rel}}$ is fixed to $10^{-3}$, while in the nonunital ansatz it is variationally optimized.
    (c) Evolution of the Bloch-vector component $x$ during the optimization shown in (b). The $(+)$ and $(-)$ labels denote the initial values that happen to converge to the two symmetry-broken steady-state sectors.
    (d) Heatmap of $|x_{\mathrm{ss}}|$ as functions of $\theta_y$ and $\theta_z$ for the Frobenius-norm cost function with $\theta_{\mathrm{dep}}=\theta_{\mathrm{rel}}=10^{-3}$ and $\epsilon=10^{-5}$. The plot is evaluated at the value of $\theta_x$ for which $|x_{\mathrm{ss}}|$ is maximal.
    (e) Variational optimization of the Frobenius-norm cost function for the nonunital and unital ansätze with $ \epsilon = 10^{-4}, \Delta = 0.5$ and $\gamma_{\mathrm{dep}}=\gamma_{\mathrm{rel}}=0.12$. In the unital ansatz, $\theta_{\mathrm{rel}}$ is fixed to $10^{-3}$.
    (f) Final energy cost as a function of depolarizing noise strength $\epsilon$, averaged over $128$ different initial conditions for $\Delta = 0.5$ for the nonunital and the unital ansatz, shows that the latter performs poorly compared to the former one.
    }
    \label{fig:Unital-nonunital-cost-comparison-plot}
\end{figure*}

Floquet systems provide a mathematical framework for understanding dynamics generated by a time-periodic Hamiltonian. In a conventional Floquet setting, a system evolves under a time‑periodic Hamiltonian $H(t + t) = H(t)$, and the full evolution over one period defines the Floquet operator
\begin{eqnarray}
    U_{F} &=& \mathcal{T}\exp[-i\int^{T}_{0} H(t) dt],
\end{eqnarray}
where $\mathcal{T}$ denotes time-ordering. Long‑time properties of the system are governed not by the instantaneous Hamiltonian but by the repeated stroboscopic action of $U_{F}$.
We define the \textit{Floquet variational ansatz} which consists of a deep circuit formed by repeating an identical parameterized layer. That is, for some variational map $\mathcal{F}(\vec{\theta})$, the circuit with $L$ layers implements
\begin{eqnarray}
    \rho_{L} &=& \underbrace{\mathcal{F}(\vec{\theta}) \circ \mathcal{F}(\vec{\theta}) \circ \cdots \mathcal{F}(\vec{\theta})}_{L \text{ times}} \rho_{0}.
\end{eqnarray}
where a single-layer ansatz is given by Eq.~\eqref{eq:Variational_ansatz_with_noise_definition}. In the limit of infinite depth $L \to \infty$, the output state is governed entirely by the fixed points of the effective channel
\begin{eqnarray}
    \rho_{\mathrm{ss}} (\vec{\theta}_{\mathrm{ss}}) &=& \mathcal{F} (\vec{\theta}_{\mathrm{ss}}) \rho_{\mathrm{ss}} (\vec{\theta}_{\mathrm{ss}}), \label{eq:Single_layer_Floquet_steady_state}
\end{eqnarray}
In other words, the steady state density matrix $\rho_{\mathrm{ss}} (\vec{\theta}_{\mathrm{ss}})$ is an eigenstate of the Floquet superoperator with eigenvalue unity. When the circuit includes only unitary gates followed by depolarizing noise, the associated Floquet superoperator becomes increasingly contractive toward the maximally mixed state $\rho_{\mathrm{ss}} = I/2$. This forces all parameter dependence to vanish, which manifests as the noise‑induced barren plateau phenomenon. Introducing nonunitary elements into this ansatz fundamentally alters this behavior. Because dissipative channels can reshape the fixed‑point structure of $\mathcal{F}(\vec{\theta})$. We show this numerically in Fig.~\ref{fig:Unital-nonunital-cost-comparison-plot}(a) and Fig.~\ref{fig:Unital-nonunital-cost-comparison-plot}(d), where we plot the magnitude of the $x$-component of the steady state density matrix as functions of the unitary angles $\theta_{y}, \theta_{z}$, for $\theta_{\mathrm{dep}} = \theta_{\mathrm{rel}} = 10^{-3}$. We find that as the nonunitary angles are nonzero, the Floquet superoperator $\mathcal{F}(\vec{\theta}_{\mathrm{ss}})$ exhibits nontrivial steady states, \textit{i.e.} $\rho_{\mathrm{ss}} \neq I/2$, which is indicated by $x_{\mathrm{ss}}$ being nonzero. If instead we consider the unitary ansatz for which $\theta_{\mathrm{dep}} = \theta_{\mathrm{rel}} = 0$, the only possible steady state is $\rho_{\mathrm{ss}} = I/2$ and $x_{\mathrm{ss}} = 0$ for all values of the unitary angles $\theta_{x}, \theta_{y}, \theta_{z}$.

\textit{Unital ansatz} --- We now discuss the effect of unital and nonunital noise on the Floquet variational ansatz (Eq.~\eqref{eq:Single_layer_Floquet_steady_state}). A quantum channel $\mathcal{E}$ is said to be unital if it preserves the identity operator, $\mathcal{E}(I) = I$, and nonunital otherwise. In our model, the depolarizing noise and the dephasing $(I+\sigma^{z})/2$ are unital. In contrast, the relaxation jump operator $\sigma^{-}$ is nonunital. To understand the effect of the nonunital noise on the optimization, we consider the case when we keep the parameter $\theta_{\mathrm{rel}}$ fixed during the optimization. In this regime, the relaxation channel acts as a static dissipative background with a fixed strength. In contrast, when $\theta_{\mathrm{rel}}$ is allowed to vary during optimization, the nonunital component becomes tunable and can actively reshape the fixed-point structure of the effective Floquet superoperator. It has been shown that noise-induced barren plateaus can be avoided if nonunital noise is present in the circuit \cite{mele2024noiseinducedshallowcircuitsabsence}. We now show that within this Floquet structure, the unital variational ansatz performs poorly compared to the nonunital one.

In Fig.~\ref{fig:Unital-nonunital-cost-comparison-plot}(b), Fig.~\ref{fig:Unital-nonunital-cost-comparison-plot}(c), Fig.~\ref{fig:Unital-nonunital-cost-comparison-plot}(e), Fig.~\ref{fig:Unital-nonunital-cost-comparison-plot}(f), we compare the behavior of the nonunital and the unital ansätze by tracking both the cost function and the corresponding order parameter $x$ over the course of optimization for the two cost functions $C_{E}[\rho]$ and $C_{F}[\rho]$. In the case of the unital ansatz, we keep the parameter $\theta_{\mathrm{rel}} = 10^{-3}$ fixed during optimization, whereas in the nonunital ansatz, we let every unitary and nonunitary parameter vary. The results show that both the unital and nonunital ansatz converge to a phase where $x_{\mathrm{ss}} \neq 0$, but the final cost value and $|x_{\mathrm{ss}}|$ are higher in the nonunital case compared to the unital one.

For the unitary Floquet ansatz, \textit{i.e} $\theta_{\mathrm{dep}} = 0$, with unital noise $\theta_{\mathrm{rel}} = 10^{-3}$, the gradients vanishes.This is how NIBPs manifest in the Floquet steady state, and clearly prevent convergence to the correct steady state for the unitary ansatz.

For the nonunitary ansatz, by contrast, we expect the Floquet solution to converge to the correct answer as $\epsilon \to 0$. Specifically, in the noiseless case $\epsilon = 0$, the Floquet ansatz can mimic Trotterization, which converges to the steady state for small time steps. Due to the stability of the driven-dissipative steady state, this solution should be perturbatively stable in $\epsilon$, implying that the exact steady state should be achievable via zero noise extrapolation~\cite{PhysRevLett.119.180509,PhysRevX.7.021050}

\section{Application: steady state electron transport in OPE-SMe}
Having validated the mechanism behind our nonunitary ansatz in an analytically tractable model, we now apply the same framework to a realistic open‑quantum system: steady‑state electron transport through an oligophenylethynylene‑sulfurmethyl (OPE-SMe) molecular junction. Our goal is to qualitatively model and interpret the experimentally observed solvent‑dependent transport reported in Ref.~\cite{TANG2020944} within a Lindblad open‑system framework that is amenable to simulation on noisy quantum hardware. In the experiment of Tang \emph{et al.}~\cite{TANG2020944}, single‑molecule junctions based on OPE-SMe were formed between two gold electrodes using the mechanically controllable break‑junction (MCBJ) technique. The electric current through the molecular junction was measured while systematically varying the solvent environment surrounding the molecule. By changing the solvent polarity using nonpolar solvents such as 1,3,5‑trimethylbenzene (TMB) and polar solvents such as tetrahydrofuran (THF) or acetonitrile (ACN), the experiment demonstrated a pronounced solvent‑induced gating effect: the single‑molecule conductance of OPE-SMe increased by nearly an order of magnitude in more polar environments. 

In our work, we model these solvent‑dependent transport regimes by treating the OPE-SMe junction as an open quantum system where the molecule is coupled to the metallic leads and driven into a nonequilibrium steady state. Within this description, the experimentally observed conductance differences between low-polarity and high-polarity solvent environments translate into different steady‑state charge currents. Rather than attempting to reproduce the detailed break‑junction dynamics, we focus on capturing the qualitative effect of the solvent on the steady state current, and on assessing whether a nonunitary variational ansatz can reliably converge to the correct steady state in the presence of realistic noise. In the next section, we describe our theoretical model for the molecular lead junction and computational methods used to obtain the Hamiltonian of OPE-SMe.

\subsection{Model for calculating the electron transport}
\begin{figure}
    \centering
    \includegraphics[width=0.99\linewidth]{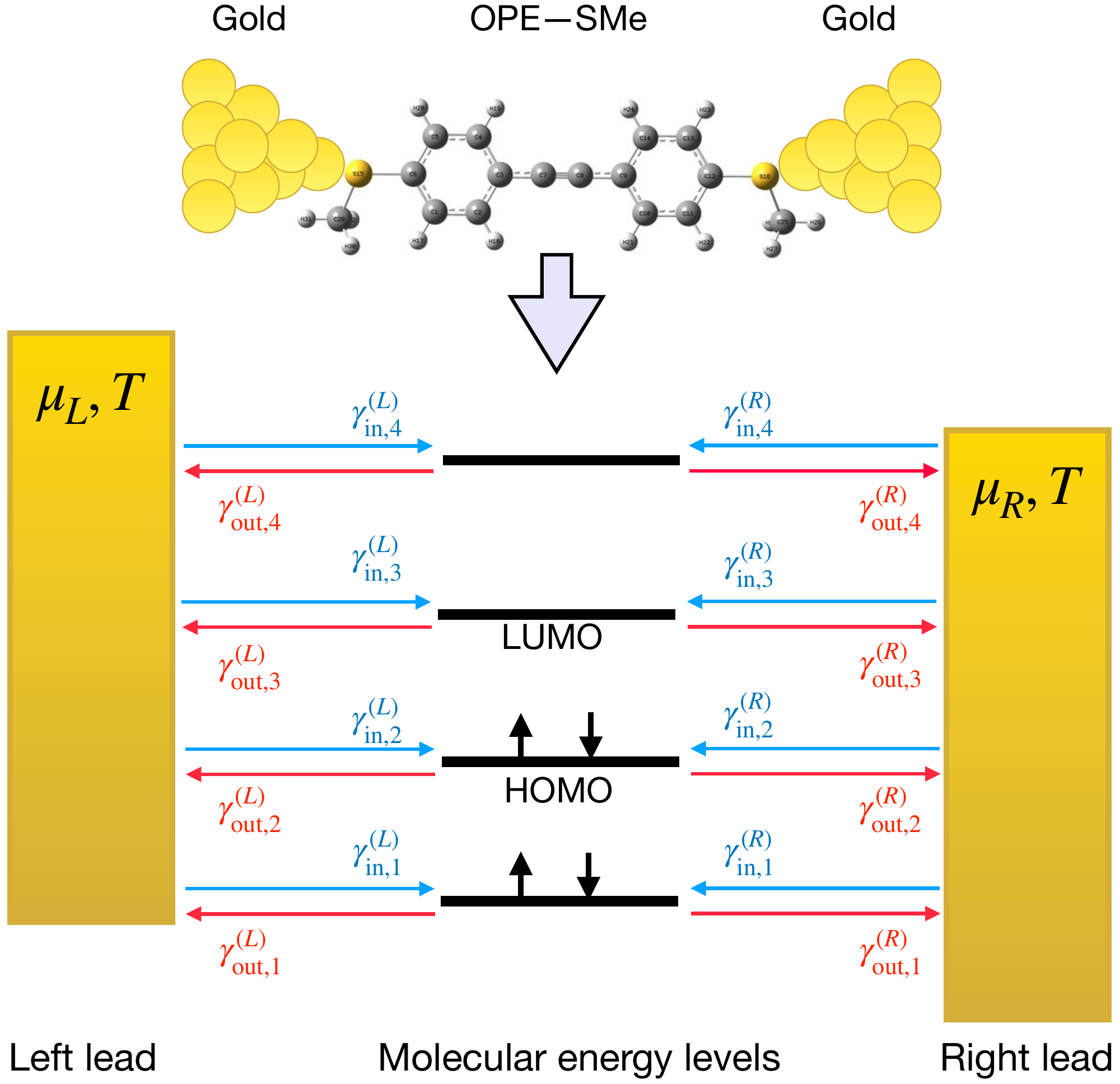}
    \caption{Schematic of electron transport model through OPE-SMe whose orbitals are coupled to the left and the right gold leads. We select four orbitals --- $(\mathrm{HOMO}-1, \mathrm{HOMO}, \mathrm{LUMO}, \mathrm{LUMO} + 3)$ --- which interacts with the left and the right lead through the processess $L_{\mathrm{in}, n} = \sqrt{\gamma_{\mathrm{in}, n}} c^{\dagger}_{n}$ (electron transfer from the lead to the molecule) and $L_{\mathrm{out}, n} = \sqrt{\gamma_{\mathrm{out}, n}} c^{\dagger}_{n}$ (electron transfer from the molecule to the leads), here $n$ represents the index of the molecular orbital. The two leads are assumed to be at chemical potential $\mu_{L}$ (left) and $\mu_{R}$ (right) and at temperature $T > 0$. The exact values of the tunneling rates used in our calculations are given in the Supplementary material (see Sec. S2)}
    \label{fig:Model_for_OPE-SME_with_metal_leads}
\end{figure}
We model the experimental setup in \cite{TANG2020944} as an open quantum system consisting of the system (Hamiltonian of the molecule + solvent if present) and the environment (the metal leads). A schematic of our model is shown in Fig.~\ref{fig:Model_for_OPE-SME_with_metal_leads}. We use first‑principles quantum‑chemistry calculations combined with an active‑space reduction to obtain an effective low‑energy Hamiltonian for the OPE-SMe molecule, whose orbitals are coupled to the left and right metallic leads, which we treat as Markovian baths. We represent the interaction between the molecule and lead using the fermionic operators $c^{\dagger}_{n} (c_{n})$, which create (annihilate) an electron in the molecular orbital $n$. The time-evolution of the whole system is governed by the Lindblad equation that consists of the Hamiltonian of the molecule, denoted by $H$, the jump operators for each orbital $n$: $L_{\mathrm{in}, n} = \sqrt{\gamma_{\mathrm{in}, n}} c^{\dagger}_{n}$ and $L_{\mathrm{out}, n} = \sqrt{\gamma_{\mathrm{out}, n}} c_{n}$. To make the simulation computationally feasible, we only consider tunneling from certain orbitals whose tunneling rates are above a threshold $\gamma_{\mathrm{thr}}$. Then the Lindblad equation for the system (molecule) density matrix can be written as
\begin{eqnarray}
\frac{d\rho}{dt}
&=& -\frac{i}{\hbar}[H,\rho] + \sum_{\gamma^{L}_{\mathrm{in}, n} > \gamma_{\mathrm{thr}}}\gamma^{L}_{\mathrm{in}, n}\Big(
c^{\dagger}_{n} \rho c_{n}
- \frac{1}{2}\{\,c_{n}c^{\dagger}_{n},\,\rho\,\}
\Big) \nonumber \\
&& + \; \sum_{\gamma^{L}_{\mathrm{out}, m} > \gamma_{\mathrm{thr}}}\gamma^{L}_{\mathrm{out}, m}\Big(
c_{m} \rho c^{\dagger}_{m}
- \frac{1}{2}\{\,c^{\dagger}_{m}c_{m},\,\rho\,\}
\Big)\nonumber \\
&& + \; \sum_{\gamma^{R}_{\mathrm{in}, n} > \gamma_{\mathrm{thr}}}\gamma^{R}_{\mathrm{in}, n}\Big(
c^{\dagger}_{n} \rho c_{n}
- \frac{1}{2}\{\,c_{n}c^{\dagger}_{n},\,\rho\,\}
\Big) \nonumber \\
&& + \; \sum_{\gamma^{R}_{\mathrm{out}, m} > \gamma_{\mathrm{thr}}}\gamma^{R}_{\mathrm{out}, m}\Big(
c_{m} \rho c^{\dagger}_{m}
- \frac{1}{2}\{\,c^{\dagger}_{m}c_{m},\,\rho\,\}
\Big)
\label{eq:Lindblad-equation-multilevel-tunneling-expression}
\end{eqnarray}
 where we have restored the reduced Planck's constant $\hbar$. The superscripts $L$ and $R$ denote the tunneling rates from the left and the right leads, respectively. 

We write the net current flowing through the left-lead + molecule + right-lead system as
\begin{eqnarray}
        I_{\mathrm{total}} &=& e \frac{d \langle \hat{N} \rangle}{dt} = I_{\mathrm{in}} + I_{\mathrm{out}} \label{eq:Total-current-expression}
\end{eqnarray}
where $\hat{N} = \sum_{k} c^{\dagger}_{k} c_{k} =  \sum_{n} (I - \sigma^{z}_{n})/2$ represents the number operator and the index $n$ runs over the orbital indices that participate in tunneling, and $e = -1.602 \times 10^{-19}$ C is the charge of electron. In the steady state, the total current vanishes: $I_{\mathrm{total}} = 0$.  Therefore $I_{\mathrm{total}}$ does not serve as an useful diagnostic for investigating the effect of the solvent on the transport. We choose the observable as $I_{\mathrm{measured}} \equiv I_{\mathrm{in, left}} - I_{\mathrm{out, left}} = I_{\mathrm{out, right}} - I_{\mathrm{in, right}}$, which represents the current which would be measured by an ammeter placed in series with the molecular junction. Using Eq.~\eqref{eq:Lindblad-equation-multilevel-tunneling-expression} in Eq.~\eqref{eq:Total-current-expression}, we obtain the following expressions (calculations in Appendix~\ref{sec:Current-expectation-value})
\begin{align}
    I_{\mathrm{measured}} &=
    |e| \left[
        \sum_{n} \gamma^{L}_{\mathrm{in}, n}
        \bigl(1 - \langle \hat{N}_{n} \rangle \bigr)
        -
        \sum_{m} \gamma^{L}_{\mathrm{out}, m}
        \langle \hat{N}_{m} \rangle
    \right], \label{eq:I_measured}
\end{align}
and equivalently,
\begin{align}
    I_{\mathrm{measured}} &=
    |e| \left[
        \sum_{n} \gamma^{R}_{\mathrm{out}, n}
        \langle \hat{N}_{n} \rangle
        -
        \sum_{m} \gamma^{R}_{\mathrm{in}, m}
        \langle \bigl(1 -\hat{N}_{m}\bigr) \rangle
    \right], \label{eq:I_measured_equivalent}
\end{align}
where $\hat{N}_{n} = (I - \sigma^{z}_{n})/2$ represents the occupancy of the $n$-th energy level of the Hamiltonian $H$, and the summations are over the energy levels which participate in tunneling. We derive the tunneling rates $\gamma^{a}_{\mathrm{in}, n}, \gamma^{a}_{\mathrm{out}, m}$ for $a=L,R$ using the Fermi golden rule (derivation in Appendix~\ref{sec:gamma_in_gamma_out_from_Fermi's_golden_rule})
\begin{eqnarray}
\gamma^{a}_{\mathrm{in}, n} &=& \frac{2 \pi}{\hbar} J^{2} \rho(\epsilon_{n}) f(\epsilon_{n}) , \\
\gamma^{a}_{\mathrm{out}, m} &=& \frac{2 \pi}{\hbar} J^{2} \rho(\epsilon_{m}) (1 - f(\epsilon_{m})) \label{eq:Lindblad-tunneling-rates}
\end{eqnarray} 
where $J^{2}$ is the tunneling matrix element between the molecule and the leads, $\rho(\epsilon_{n})$ is the density of states of the leads, $f(\epsilon_{n})$ is the Fermi-Dirac distribution $f(\epsilon_{n}) = (e^{(\epsilon_{n} - \mu) / (k_{B} T)} + 1)^{-1}$, $\mu$ is the chemical potential, $T$ is the temperature in absolute scale, and $k_{B} = 8.617 \times 10^{-5}$ eV / K is the Boltzmann constant. In the next section, we describe the variational scheme used to determine the steady state current $I_{\mathrm{in}}$ through the molecule.
\begin{figure}
    \centering
    \includegraphics[width=\linewidth]{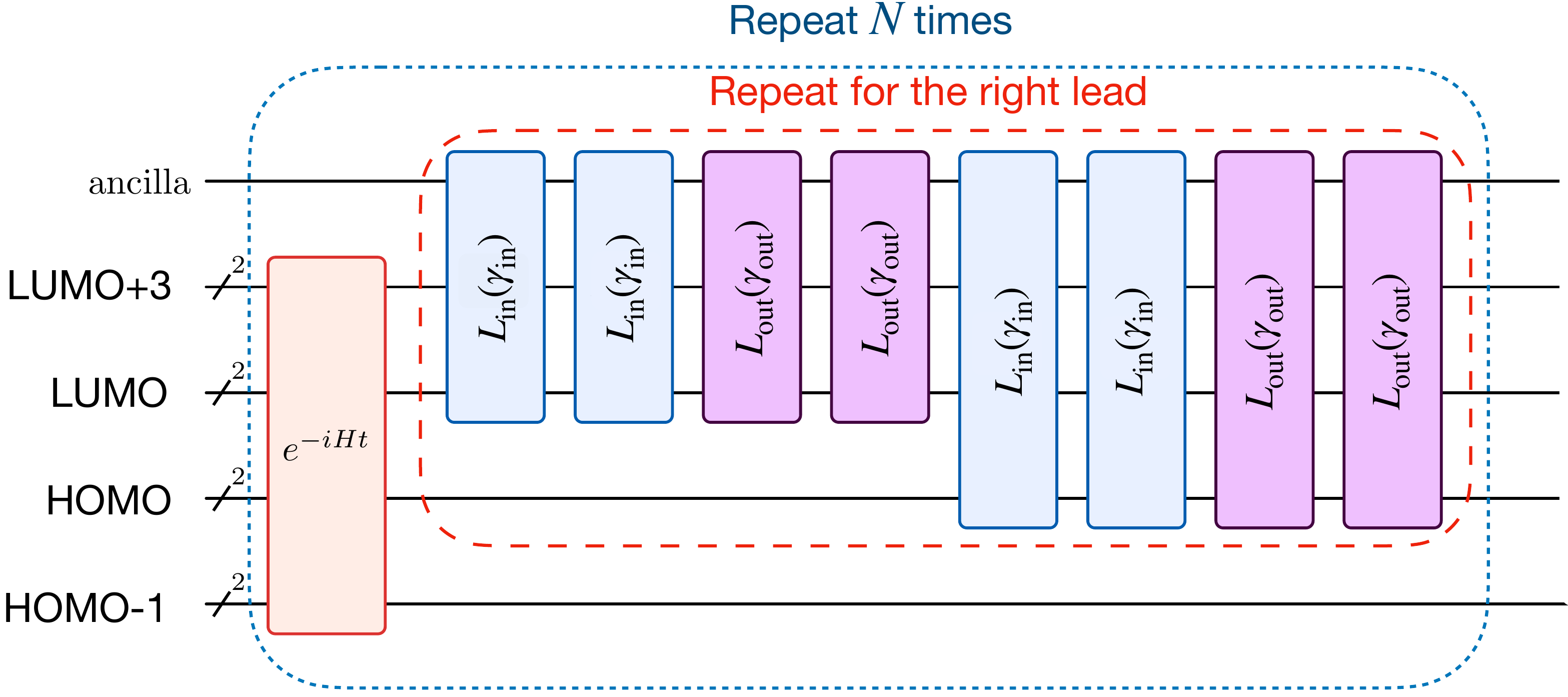}
    \caption{Trotterized Lindbladian time evolution of the molecular transport model implemented using one ancilla qubit. The coherent time evolution generated by the molecular Hamiltonian acts only on the system qubits, while dissipative tunneling processes are implemented via ancilla-assisted nonunitary gates. For simplicity of the figure, we assume that only the frontier orbitals---HOMO and LUMO---have nonzero tunneling rates to and from the left and right metallic leads. After the Jordan--Wigner transformation, each fermionic orbital is represented by two qubits, which is indicated schematically by the symbol $/^{2}$ on each horizontal wire. The gates $L_{\mathrm{in}}(\gamma_{\mathrm{in}})$ and $L_{\mathrm{out}}(\gamma_{\mathrm{out}})$ implement the Lindblad jump operators $\sqrt{\gamma_{\mathrm{in}}}\,c_n^{\dagger}$ and $\sqrt{\gamma_{\mathrm{out}}}\,c_n$, respectively, where $n=\text{HOMO}$ or $\text{LUMO}$. Each nonunitary gate is drawn twice on the same wire to emphasize the two-qubit representation of each orbital. The explicit circuit constructions for $L_{\mathrm{in}}(\gamma_{\mathrm{in}})$ and $L_{\mathrm{out}}(\gamma_{\mathrm{out}})$ are shown in Fig.~~\ref{fig:L_in_L_out_operator_circuit_diagrams}. The block enclosed by the red dotted box represents tunneling processes associated with one lead and is repeated for the opposite lead. In this schematic, the same orbitals are assumed to couple to both leads; in general, tunneling from and to the left and right leads must be treated separately. The full Trotterized circuit, including the Hamiltonian and dissipative steps, is repeated $N$ times with time step $\delta t = T/N$, where $T$ is the total evolution time.}
    \label{fig:Circuit-implementation-Lindblad}
\end{figure}

\subsection{Variational ansatz for steady state current calculation} \label{subsec:Variational ansatz for steady state current calculation}
We now formulate the variational ansatz to determine the non-equilibrium steady state of the OPE-SMe transport model and to evaluate the associated steady-state charge current. The central objective is to approximate the fixed point of the Lindblad generator in Eq.~~\eqref{eq:Lindblad-equation-multilevel-tunneling-expression} using a parameterized family of completely positive trace-preserving (CPTP) maps that can be implemented efficiently on near-term quantum hardware. The exact steady state $\rho_{\mathrm{ss}}$ of the system is defined by the condition
\begin{eqnarray}
\mathcal{L}\rho_{\mathrm{ss}} = 0 ,
\end{eqnarray}
where $\mathcal{L}$ is the full Lindblad superoperator incorporating both unitary evolution under the molecular Hamiltonian and nonunitary injection and extraction processes due to the metallic leads. To target the steady state, we minimize the Frobenius norm cost function $C[\rho] = \mathrm{Tr}[(\mathcal{L} \rho)^{2}]$, which vanishes when the system reaches the steady state: $\rho \equiv \rho_{\mathrm{ss}}$. We define the nonunitary variational ansatz constructed directly from the physical generators of the problem. Each variational layer is defined as
\begin{eqnarray}
    \mathcal{U}(\theta_{1}, \theta_{2}, \theta_{3}) &\equiv & e^{\mathcal{H}\theta_{1}} e^{\mathcal{L}_{\mathrm{in}} \theta_{2}} e^{\mathcal{L}_{\mathrm{out}} \theta_{3}} \label{eq:Variational-ansatz-OPESMe}
\end{eqnarray}
where $\mathcal{H}[\rho] = -i[H,\rho]$ generates unitary evolution under the molecular Hamiltonian, and $\mathcal{L}_{\mathrm{in}}$ and $\mathcal{L}_{\mathrm{out}}$ denote the Lindblad dissipators associated with electron injection from and extraction into the leads, respectively. For the circuit implementation of this ansatz, we replace $\gamma_{\mathrm{in}} dt$ by $\theta_{2}$ and $\gamma_{\mathrm{out}} dt$ by $\theta_{3}$ in the Trotterized circuit in Fig.~\ref{fig:Circuit-implementation-Lindblad}. Note that every $\mathcal{L}_{\mathrm{in}}$ ($\mathcal{L}_{\mathrm{out}}$) share the same parameter $\theta_{2}$ $ (\theta_{3})$. We use gradient descent [Eq.~\eqref{eq:Gradient-descent-rule-for-parameter-update}] to update the parameters during optimization. The standard parameter-shift rule applies only to expectation-value cost functions which are of the form $\mathrm{Tr}[O \rho]$, and to ans\"atze consisting of unitary gates \cite{PhysRevA.98.032309, PhysRevA.99.032331}. Because our ansatz contains nonunitary elements, we cannot use the parameter shift rule to evaluate the gradient of the cost function. Instead, we use the finite difference
\begin{eqnarray}
    \frac{\partial C}{\partial \theta} &\approx & \frac{C(\theta + \delta/2) - C(\theta - \delta/2)}{\delta}
\end{eqnarray}
where $\delta$ is a small positive quantity.

In order to benchmark the variational results, we also simulate the full Lindbladian dynamics using Trotterization using a quantum circuit.
The exact solution of Eq.~\eqref{eq:Lindblad-equation-multilevel-tunneling-expression} is given by $\rho(t) = e^{\mathcal{L} t}\rho(0)$, where the Liouvillian $\mathcal{L}$ contains both the coherent Hamiltonian contribution and the dissipative jump processes associated with electron injection and extraction. Because the individual terms in $\mathcal{L}$ do not commute, we approximate the evolution operator over a small time step $\delta t$ using a first-order Suzuki-Trotter decomposition
\begin{eqnarray}
    e^{\mathcal{L}\delta t}
    \approx
    e^{\mathcal{H}\delta t}
    \prod_{\alpha} e^{\mathcal{L}_{\alpha}\delta t} \label{eq:Suzuki_Trotter_decomposition}
\end{eqnarray}
where $\mathcal{H}[\rho]=-i[H,\rho]$ generates unitary evolution under the molecular Hamiltonian and $\mathcal{L}_{\alpha}$ denotes the individual Lindblad dissipators corresponding to tunneling processes between the molecule and the leads. The unitary component $e^{\mathcal{H}\delta t}$ is implemented directly using standard Hamiltonian simulation techniques after mapping the fermionic Hamiltonian to qubits via the Jordan-Wigner transformation. The nonunitary terms $e^{\mathcal{L}_{\alpha}\delta t}$ are realized using ancilla-assisted circuits that implement amplitude-damping-type channels corresponding to the jump operators $L_{\mathrm{in}}=\sqrt{\gamma_{\mathrm{in}}}c^\dagger$ and $L_{\mathrm{out}}=\sqrt{\gamma_{\mathrm{out}}}c$. We derive the exact quantum circuits implementing these jump operators in Appendix ~\ref{sec:Jump-operators-using-ancilla-qubits}. Each dissipative step couples the system qubits that participate in tunneling to and from the leads to an ancilla qubit, and then applies a controlled rotation whose angle is determined by the tunneling rate for the orbital $\gamma_{\alpha}$ and the time-step of the Trotterization scheme $\delta t$, and then the ancilla is reset, thereby realizing a nonunitary process for the system. By repeating this Trotterized circuit for $N$ time steps with $t=N\delta t$, the system is evolved toward its nonequilibrium steady state. The quantum circuit implementing the time-evolution using Eq.~\eqref{eq:Suzuki_Trotter_decomposition} is shown in Fig. ~\ref{fig:Circuit-implementation-Lindblad}. To accommodate the nonunitary time-evolution, we modified the existing TrotterQRTE class in Qiskit \cite{qiskit_trotter_qrte_tutorial}, where each time-step is augmented by an ancilla-assisted implementation of the jump operators (Fig. ~\ref{fig:Circuit-implementation-Lindblad}). We then measure the expectation value of the observable in Eq.~\eqref{eq:I_measured} after each time step, which allows us to extract the long-time steady-state currents directly from the circuit.
In the next section, we describe the quantum mechanics/molecular mechanics (QM/MM) methodology used to construct an effective low-energy Hamiltonian for the OPE-SMe molecule, the calculation of the density of states together with the first principles calculation to approximate $\rho(\epsilon)$ and $J$.

\section{Vacuum and Solvated OPE-SMe Models}

To obtain the molecular Hamiltonian of OPE-SMe in different solvent environments, we performed polarizable QM/MM calculations using LICHEM~\cite{lichem, gokcan2019lichem}, interfacing Gaussian~16~\cite{gaussian16} (QM) with TINKER~\cite{tinker} (MM, AMOEBA polarizable force field). The QM region comprised all 32 OPE-SMe atoms treated at the B3LYP/6-31+G(d,p) level of theory, while the MM region consisted of 500 explicit THF molecules in a periodic cubic box. Two representative solvation snapshots were selected from an NPT molecular dynamics trajectory: O-Away and O-Toward environments. A strongly O-away frame (frame 92, collective orientation score $S = -0.410$) and a near-isotropic frame (frame 929, $S = -0.005$) were selected. These two configurations were chosen to maximize the electrostatic contrast between solvation environments, and the resulting polarized electron densities were used to construct environment-specific Hamiltonians for the transport calculations.

\begin{figure*}
    \centering
    \includegraphics[width=0.85\textwidth]{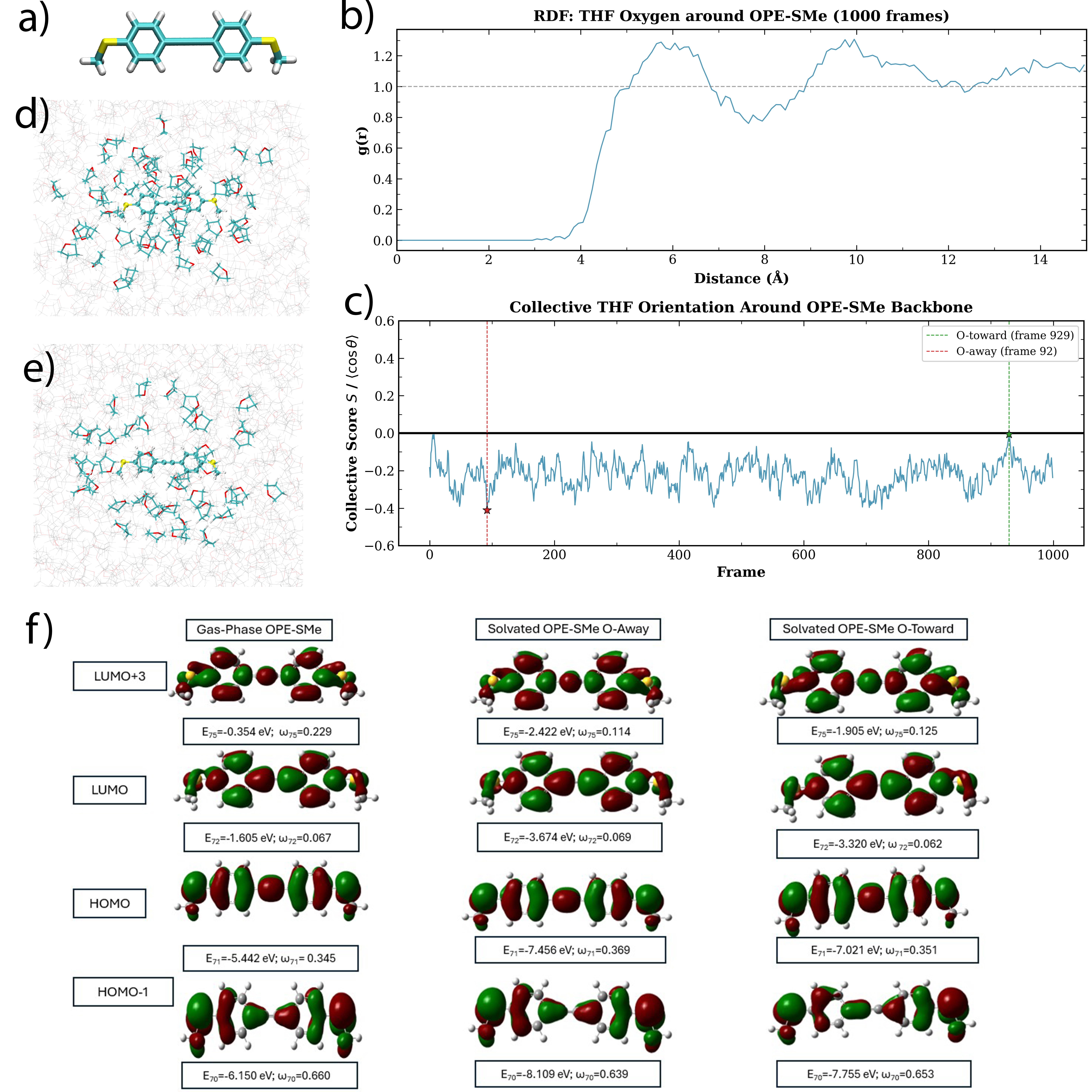}
    \caption{a) Gas-phase OPE-SMe geometry at the B3LYP/6-31+G(d,p) level. b) Radial distribution function $g(r)$ for THF oxygen atoms around the OPE-SMe solute, averaged over 1000 MD frames. The first maximum appears at $r \approx 5.75$~\AA\ ($g \approx 1.29$) with a shallow minimum at $r \approx 7.65$~\AA\ ($g \approx 0.76$, running coordination number $n \approx 0.025$), confirming a weakly structured first solvation shell. (c) Time evolution of the collective THF orientation score $S(t) = \langle \cos\theta \rangle$ over the 1000-frame (100 ps) trajectory. The score remains negative throughout, indicating a persistent O-away preference across the entire simulation. Stars mark extreme snapshots: frame 92 (O-away, $S = -0.410$, red dashed line and red star) and frame 929 (near-isotropic, $S = -0.005$, green dashed line and green star), selected as representative configurations for subsequent QM/MM single-point calculations. (d) O-Away solvated configuration (MD frame~92, $S = -0.410$), with THF oxygen atoms (red) predominantly oriented away from the OPE-SMe $\pi$-scaffold. Color scheme: sulfur = yellow, carbon = cyan, oxygen = red. (e) Near-isotropic (O-Toward) solvated configuration (MD frame~929, $S = -0.005$), showing a more balanced THF orientational distribution around the backbone. (f) Frontier molecular orbitals (HOMO$-$1, HOMO, LUMO, and LUMO$+$3) for OPE-SMe in three environments: gas phase (left), solvated O-Away (center), and solvated O-Toward (right), obtained from polarizable QM/MM single-point calculations at the B3LYP/6-31+G(d,p) level. Orbital energies $E$ and sulfur-projected weights $\omega$ are indicated for each MO. Solvation stabilizes all frontier orbitals, with the O-Away environment producing a larger shift than the O-Toward environment.}
    \label{fig:computational-methods}
\end{figure*}

Figure 10(f) shows the frontier molecular orbitals (HOMO$-$1, HOMO, LUMO, and LUMO$+$3) and their sulfur-projected weights across the three environments (gas phase, O-Away, and O-Toward). Solvation stabilizes all frontier orbitals, with the O-Away environment producing a larger energy shift than the O-Toward environment, directly influencing the molecule-electrode coupling strengths and the resulting steady-state transport properties. For a detailed description of the force-field parameterization, MD simulation protocol, solvent orientation analysis, and active-space construction, we refer the reader to Appendix E.

\subsection{Results: steady-state currents in OPE-SMe}
We now present the results of the molecular transport: the convergence of the nonunitary variational ansatz to the nonequilibrium steady state (NESS) of the OPE-SMe junction, and the resulting steady-state charge currents $I_{\mathrm{measured}}$ across different molecular environments. We calculate the tunneling rates using Eq.~\eqref{eq:Lindblad-tunneling-rates} at temperature $T = 300$ K, chemical potentials of the left and the right leads at $\mu_{\mathrm{left}} = -3.29$ eV and $\mu_{\mathrm{right}} = -3.75$ eV, and using the energies listed in Supplementary material (see Sec. S1). We have chosen the chemical potentials for the leads such that they lie just above and below the midpoint of the HOMO and LUMO energies of OPE-SMe in gas phase. We initialize the system in the state $|\uparrow \uparrow 0 0\rangle \equiv |00110011\rangle$ \footnote{To see the equivalence of these two representations of the same state, note that Jordan Wigner transformation in Qiskit maps the orbitals in fermionic representation $(69, 70, 71, 72)$ to the qubits representation $(69\alpha, 70\alpha, 71 \alpha, 72\alpha, 69\beta, 70\beta, 71\beta, 72\beta)$, where $\alpha$ and $\beta$ represents the two spins of the electron [see \href{https://qiskit-community.github.io/qiskit-nature/stubs/qiskit_nature.second_q.mappers.InterleavedQubitMapper.html}{documentation from Qiskit website on how this mapping works}]. Moreover, while representing quantum states, Qiskit follows the convention of counting from right to left---the rightmost qubit is the zeroth qubit. Therefore, the initial state $|$HOMO$-1$, HOMO, LUMO, LUMO$+1$ $\rangle = |\uparrow \uparrow 00\rangle$ gets mapped to the state $|00110011\rangle$.}, which represents the state where the orbitals are filled up to the HOMO. Note that the number of jump operators doubles after they are mapped from the fermionic to the qubit representation.

\begin{figure}
    \centering
    \includegraphics[width=0.9\linewidth]{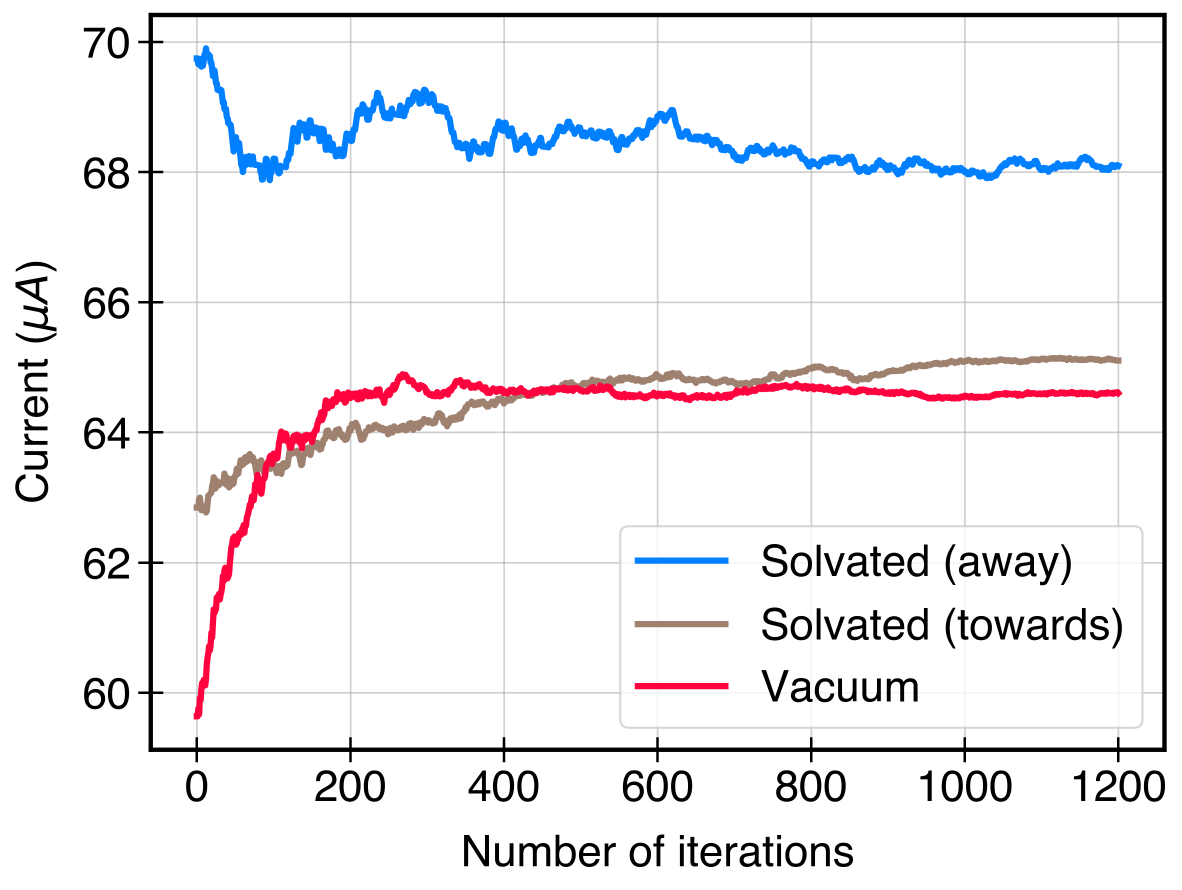}
    \caption{The observable $I_{\mathrm{measured}}$ (Eq.~\eqref{eq:I_measured}) a function of the number of iterations of the single layer variational ansatz obtained from Qiskit simulations on the FakeTorino backend, which mimics the real device.}
    \label{fig:OPES--Me-variational-currents-FakeTorino}
\end{figure}

\textit{Simulation on a fake backend noise model}: We first simulate the steady state transport on a FakeTorino backend, which emulates a $133$-qubit IBM quantum device using a system snapshot that includes realistic hardware characteristics such as nontrivial connectivity of the qubits, supported basis gates, qubit relaxation and dephasing times $(T_{1}, T_{2})$, and gate and readout error rates \cite{ibm_fake_provider_api}. Fig. ~\ref{fig:OPES--Me-variational-currents-FakeTorino} shows the current $I_{\mathrm{measured}}$ as a function of iteration number for the three environments considered (vacuum and two solvated configurations). In all cases, the current convergence to a stable plateau indicates that the variational procedure successfully reaches a nonequilibrium steady state despite the presence of realistic noise. The steady-state current values reached by the algorithm differ across the three environments, demonstrating that the variational circuit remains sensitive to environment-dependent changes in the molecular Hamiltonian and tunneling rates. This sensitivity persists despite the accumulation of depolarizing noise, decoherence, and readout errors throughout the optimization loop. This suggests that the converged steady states reflect physically meaningful steady states rather than noise-induced saturation.

\begin{figure}
    \centering
    \includegraphics[width=0.99\linewidth]{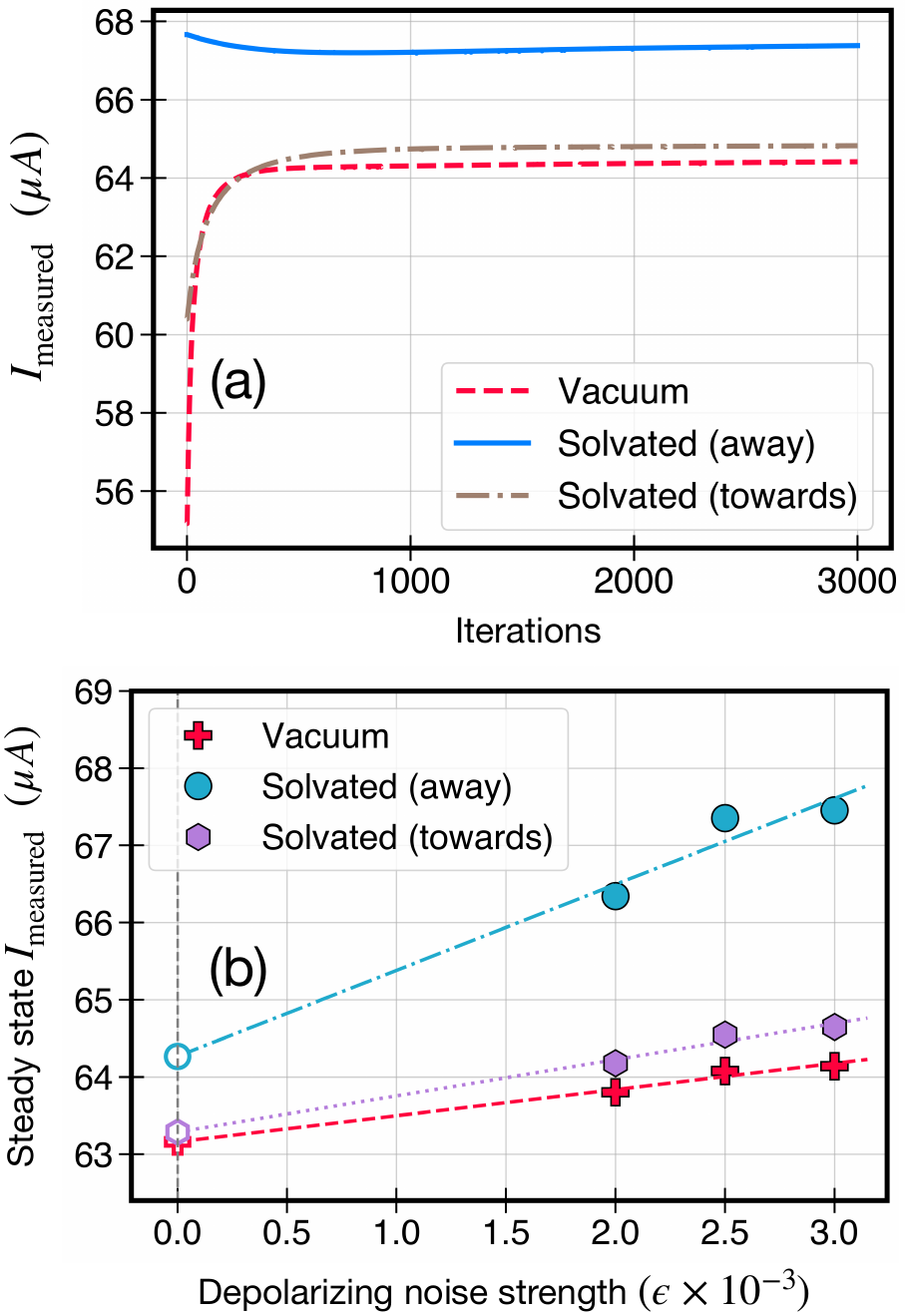}
    \caption{(a) Charge current as a function of number of iterations of the variational algorithm, obtained from Qiskit simulations with a depolarizing noise model with strength $\epsilon = 2.5 \times 10^{-3}$. The currents are calculated using the three-parameter variational ansatz in Eq.~\ref{eq:Variational-ansatz-OPESMe}. (b) Corresponding steady-state charge currents for the three environments, extrapolated to the zero-noise limit ($\epsilon \rightarrow 0$)}
    \label{fig:OPE-SMe-depolarizing-noise-simulation-zero-noise-extrapolation}
\end{figure}

\begin{figure}
    \centering
    \includegraphics[width=0.84\linewidth]{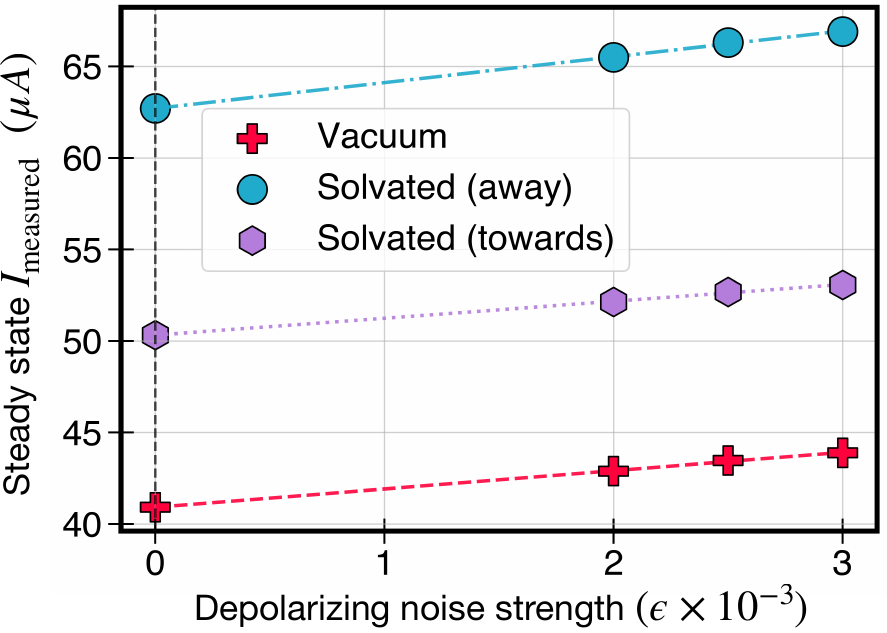}
    \caption{Zero-noise–extrapolated steady-state currents obtained via Trotterized time evolution (see Fig. ~\ref{fig:Circuit-implementation-Lindblad}), simulated in Qiskit with a depolarizing noise model.}
    \label{fig:OPE-SMe-Trotterization-zero-noise-extrapolation}
\end{figure}

\emph{Simulation with a depolarizing noise model}: We now run controlled depolarizing-noise simulations of the nonunitary variational algorithm. To validate the variational steady-state currents, we benchmark the variational current against the exact time-evolution using Trotterization, which approximates $\rho(t)=e^{\mathcal{L}t}\rho(0)$ through repeated application of coherent and dissipative Trotter steps. But, the direct comparison between the Trotterized and the variational steady state currents at the same noise strength $\epsilon$ is not meaningful, because these two circuits have very different circuit depths and gate counts---the Trotterized circuit is much deeper, since every time step contains Hamiltonian evolution plus many jump blocks, whereas the variational circuit is much shallower by construction. Therefore, even if both are approximating the same physical steady state, the Trotter circuit will accumulate much more noise at fixed depolarizing noise strength $\epsilon$. We therefore compare the zero noise extrapolated steady state incoming currents obtained from the variational algorithm with the ones obtained using Trotterized time-evolution. We optimize the Frobenius-norm cost function defined in Eq.~\eqref{eq:Frobenius-norm-cost-function} using the QM/MM-derived Hamiltonian for OPE--SMe and the Lindblad jump operators $L_{\mathrm{in}}$ and $L_{\mathrm{out}}$. The steady state is approximated with the single-layer variational quantum circuit introduced in Eq.~\eqref{eq:Variational-ansatz-OPESMe}, which contains three variational parameters, $\theta_{1}$, $\theta_{2}$, and $\theta_{3}$. At each variational iteration, we evaluate the measured current observable $I_{\mathrm{measured}}$. The resulting steady-state currents are shown in Figs.~\ref{fig:OPE-SMe-depolarizing-noise-simulation-zero-noise-extrapolation} and~\ref{fig:OPE-SMe-Trotterization-zero-noise-extrapolation}.

The variational ansatz reproduces the expected qualitative ordering of the steady-state currents,
\begin{eqnarray}
I_{\mathrm{Solvated\;away}} > I_{\mathrm{Solvated\;towards}} > I_{\mathrm{Vacuum}} .
\end{eqnarray}
However, the single-layer ansatz does not quantitatively reproduce the steady-state currents obtained from Trotterized Lindblad time evolution with zero-noise extrapolation. This discrepancy is likely due to the limited expressibility of the single-layer ansatz. We expect that increasing the number of variational layers will improve convergence toward the exact steady state and yield better quantitative agreement with the Trotterized zero-noise-extrapolated results.
 
The observed qualitative agreement under zero-noise extrapolation between the variational and the Trotterized steady state currents therefore supports our claim --- nonunitary variational ans\"atze provide a practical route to simulating nonequilibrium steady states on noisy hardware.

\section{Conclusion}
In this work, we showed that noise-induced barren plateaus in variational quantum algorithms can be mitigated by using a non-unitary ansatz. We first use an all-to-all interacting dissipative Ising model, which shows two phases---paramagnetic and ferromagnetic. We showed explicitly that a nonunitary variational ansatz remains trainable under depolarizing noise and converges reliably to the correct symmetry‑broken steady state. In contrast, purely unitary ans\"atze exhibit noise‑induced barren plateaus and fail to reproduce the ferromagnetic phase. By introducing a Floquet‑type ansatz in which a single parameterized layer is repeated, we reduced the deep variational circuit to an effective quantum channel and identified the channel fixed points as the key objects governing both trainability and convergence. This channel perspective clarifies how unital noise drives variational circuits toward trivial fixed points, while tunable dissipative elements restore nontrivial steady states and prevent gradient collapse. Although the Floquet ansatz does not reach the same optimal cost value as the variational ansatz with independent parameters in each layer, it plays a crucial conceptual role in our analysis. By forcing every layer of the circuit to share the same parameters, the high‑dimensional optimization problem reduces to the study of a single quantum channel.

We further demonstrated that these ideas extend beyond analytically solvable toy models by applying the nonunitary variational framework to a realistic molecular‑transport problem: steady‑state electron flow through OPE-SMe. Using Hamiltonians and tunneling rates derived from first‑principles QM/MM calculations and implementing dissipative processes via ancilla‑assisted circuits, we showed that the variational ansatz captures the correct qualitative ordering of steady‑state currents across different molecular environments. Importantly, zero‑noise extrapolated results obtained from the variational algorithm are consistent with those from Trotterized Lindblad time evolution, despite the substantially reduced circuit depth of the variational approach.

Overall, our results establish nonunitary variational ans\"atze as a scalable and physically grounded strategy for simulating open‑quantum‑system steady states on NISQ hardware. By embedding controlled dissipation directly into the circuit, one can bypass a fundamental limitation of noisy unitary variational algorithms and access regimes that would otherwise be untrainable. Beyond quantum transport, this framework is broadly applicable to driven‑dissipative many‑body systems, non‑equilibrium phase transitions, and quantum algorithms where steady states, rather than ground states, are the central objects of interest.

\section{Acknowledgments}
This work was performed with support from the National Science Foundation (NSF) through Grant No. OSI-222872, National Institutes of Health through award No. R35GM151951, and DARPA through Award No. HR0011233002. The authors thank the High Performance Computing facility at the University of Texas at Dallas (HPC@UTD) for providing computational resources.

\section{Data availability}
The data that support the findings of this article are openly
available in \href{https://zenodo.org/records/20398248}{10.5281/zenodo.20398248}.

\appendix
\section{Quantum circuits implementing the jump operators} \label{sec:Jump-operators-using-ancilla-qubits}
In this section, we describe the implementation of jump operators in our model using ancilla qubits. The two jump operators are: $L_{\mathrm{in}} = \sqrt{\gamma_{\mathrm{in}}} c^{\dagger}_{n}, L_{\mathrm{out}} = \sqrt{\gamma_{\mathrm{out}}} c_{n}$ which creates and annihilates an electron in the energy level $n$ of the molecule. To realize these using quantum circuits, we first transform them into the spin language using the Jordan-Wigner transformation
\begin{eqnarray}
    c^{\dagger}_{n} &=& \frac{1}{2} (\sigma^{x}_{n} - i \sigma^{y}_{n}) \bigg( \prod^{L}_{k < n} \sigma^{z}_{k} \bigg), \label{eq:c_dagger_LUMO_multiqubit_Jordan_Wigner}\\
    c_{n} &=& \frac{1}{2} (\sigma^{x}_{n} + i \sigma^{y}_{n})\bigg( \prod^{L}_{k < n} \sigma^{z}_{k} \bigg),
\end{eqnarray}
where $\sigma^{z}_{k}$ represents the Pauli matrix $\sigma^{z} = \begin{bmatrix}
    1 & 0\\
    0 & -1
\end{bmatrix}$ acting on qubit $k$.
We begin by describing the implementation of the injection jump operator
$L_{\mathrm{in}}$ using a single ancilla qubit. As a first step, we consider
the realization of the lowering operator $\sigma^{-}_{n}$ via an effective
circuit involving only the system qubit $n$ and one ancilla qubit. In this
simplified setting, the Jordan-Wigner string factor
$\prod_{k<n} \sigma^{z}_{k}$ is absent and will be incorporated subsequently.

\begin{figure}[H]
    \centering
    \includegraphics[width=\linewidth]{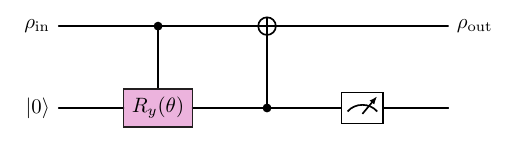}
    \caption{Circuit implementation of the $L_{\mathrm{out}}$ using a single ancilla qubit.}
    \label{fig:sigma-minus-implementation}
\end{figure}
The effective two-qubit $L_{\mathrm{out}}$ operator can be implemented using controlled $R_{y}$ and a controlled not gate \cite{Nielsen_Chuang_2010} as shown in Fig.~\ref{fig:sigma-minus-implementation}. We now show this explicitly by construction. Let $U$ represents the unitary operation that takes the input $\rho_{\mathrm{in}} \otimes |0\rangle \langle 0|$ and output density matrix $\rho_{\mathrm{out}}$ in Fig.~\ref{fig:sigma-minus-implementation}. Then we can write \cite{Nielsen_Chuang_2010}
\begin{eqnarray}
    \rho_{\mathrm{out}} &=& \mathrm{Tr}_{\rm ancilla} \bigg[U (\rho_{\mathrm{in}} \otimes |0\rangle \langle 0|) U^{\dagger}\bigg] \label{eq:Final-density-matrix-formula}
\end{eqnarray}
Let the initial density matrix be of the form
\begin{eqnarray}
\rho(0) \equiv \rho_{\mathrm{in}} = \begin{bmatrix}
    p_{0} & c \\
    c^{*} & p_{1}
\end{bmatrix},\qquad p_{0} + p_{1} = 1
\end{eqnarray}
We now show that the following unitary $U$ implements the $L_{\mathrm{out}}$ operation (see Exercise $8.20$ in Ref. \cite{Nielsen_Chuang_2010})
\begin{eqnarray}
    U = \mathrm{CNOT}\cdot \mathrm{CR}_{y}(2\theta), \;
    U^{\dagger} = \mathrm{CR}_{y}(-2\theta) \cdot \mathrm{CNOT}^{\dagger}
\end{eqnarray}
where the angle $\theta$ should be expressed in terms of the damping rate $\gamma_{\mathrm{out}}$.
The controlled gates are given by
\begin{eqnarray}
    \mathrm{CR}_{y}(2\theta) &=&  
|0\rangle \langle 0| \otimes I + |1\rangle \langle 1| \otimes R_{y}(\theta) ,
\end{eqnarray}
and
\begin{eqnarray}
    \mathrm{CNOT} &=& I \otimes |0\rangle \langle 0| + \sigma^{x} \otimes |1\rangle \langle 1| .
\end{eqnarray}
Let us now calculate the expression inside the trace in Eq.~\eqref{eq:Final-density-matrix-formula} with $U = \mathrm{CNOT} \cdot \mathrm{CR}_{y}(2\theta)$, which is
\begin{eqnarray}
 && U (\rho_{i} \otimes |0\rangle \langle 0|) U^{\dagger}\\
   &=& \begin{bmatrix}
        p_{0} & c\sin\theta & c\cos\theta & 0\\
        c^{*} \sin\theta & p_{1}\sin^{2}\theta & p_{1}\cos\theta\ \sin\theta & 0\\
        c^{*}\cos\theta & p_{1}\cos\theta \sin\theta & p_{1}\cos^{2}\theta & 0\\
        0 & 0 & 0 & 0
    \end{bmatrix}
\end{eqnarray}
Tracing over the first qubit (ancilla), we obtain
\begin{eqnarray}
    \rho_{\mathrm{out}} &=& \begin{bmatrix}
        p_{0}+p_{1}\sin^{2}\theta & c\cos\theta\\
        c^{*}\cos\theta & p_{1}\cos^{2}\theta
    \end{bmatrix} . \label{eq:Final_matrix_after_application_of_U}
\end{eqnarray}
Now, in order to express the rotation angle $\theta$ in terms of the tunneling rate $\gamma_{\mathrm{out}}$, we consider the time evolution of the density matrix under the Liouvillian with only the jump operator $\sqrt{\gamma_{\mathrm{out}}} c$ for an infinitesimal time period $dt$:
\begin{eqnarray}
    \rho(dt) &=& e^{\mathcal{L}_{\mathrm{out}}dt} \rho(0) ,\\
    &\approx & \rho(0)+dt\Dot{\rho} ,  \\
    &=& \begin{bmatrix}
        p_{0}+dt\gamma_{\rm out}p_{1} & c -dt\frac{\gamma_{\rm out}c}{2} \\
        c^{*} - dt\frac{\gamma_{\rm out}c^{*}}{2} & p_{1}-dt\gamma_{\rm out}p_{1}
    \end{bmatrix} ,\label{eq:Density-matrix-after-ancilla}
\end{eqnarray}
where we have calculated $\Dot{\rho}$ with the jump operator $L_{\mathrm{out}} = \sqrt{\gamma_{\mathrm{out}}}(\sigma^{x}+i\sigma^{y})/2$ using
\begin{eqnarray}
    \Dot{\rho} &=& \gamma_{\rm out}(L_{\rm out}\rho L^{\dagger}_{\rm out}-\frac{1}{2} \{L^{\dagger}_{\rm out} L_{\rm out},\rho \}) \\
    &=& \begin{bmatrix}
    \gamma_{\rm out}p_{1} & -\frac{\gamma_{\rm out}c}{2} \\
    -\frac{\gamma_{\rm out}c^{*}}{2} & -\gamma_{\rm out}p_{1}
    \end{bmatrix} \label{eq:Final_matrix_after_application_of_L_out}
\end{eqnarray}
Comparing the matrices Eq.~\eqref{eq:Final_matrix_after_application_of_U} and Eq.~\eqref{eq:Density-matrix-after-ancilla}, we obtain
\begin{eqnarray}
    \sin^{2}\theta &=& dt \gamma_{\rm out}\\
    \theta &=& \arcsin(\sqrt{dt \gamma_{\rm out}}) \label{eq:L_out-rotation-angle-solution-equation}
\end{eqnarray}
Thus the infinitesimal evolution of the density matrix under the jump operator $L_{\mathrm{out}}$ can be implemented using the quantum circuit shown in Fig.~\ref{fig:sigma-minus-implementation} if we choose the rotation angle according to Eq.~\ref{eq:L_out-rotation-angle-solution-equation}.

\begin{figure}[H]
    \centering
    \includegraphics[width=\linewidth]{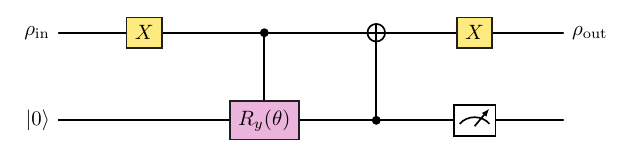}
    \caption{Implementation of $L_{\mathrm{in}}$ using ancilla qubit.}
    \label{fig:sigma-plus-implementation}
\end{figure}

Now we show how to implement $L_{\mathrm{in}}$. In this case, the unitary operator is
\begin{eqnarray}
  U &=&  (\sigma^{x} \otimes I)\cdot \mathrm{CNOT} \cdot \mathrm{CR}_{y}(2\theta) \cdot(\sigma^{x}\otimes I)
\end{eqnarray}
and the corresponding circuit is shown in Fig.~\ref{fig:sigma-plus-implementation}.
Using Eq.~\eqref{eq:Final-density-matrix-formula}, we obtain
\begin{eqnarray}
    && U (\rho_{i} \otimes |0\rangle \langle 0|) U^{\dagger} \\  & = & \begin{bmatrix}
        p_{0}\cos^{2}\theta & 0 & c\cos\theta & p_{0}\cos\theta \sin\theta\\
        0 & 0 & 0 & 0\\
        c^{*}\cos\theta & 0 & p_{1} & c^{*}\sin\theta\\
        p_{0}\cos\theta \sin\theta & 0 & c\sin\theta & p_{0}\sin^{2}\theta
    \end{bmatrix}
\end{eqnarray}
Tracing over the first qubit (ancilla)
\begin{eqnarray}
    \rho_{\mathrm{out}} &=& \begin{bmatrix}
        p_{0}\cos^{2}\theta & c\cos\theta\\
        c^{*}\cos\theta & p_{1} + p_{0}\sin^{2}\theta
    \end{bmatrix} \label{eq:Density_matrix_after_applicaion_of_U_ancilla}
\end{eqnarray}
Following the similar approach as in $L_{\mathrm{out}}$, for $L_{\rm in} = \sqrt{\gamma_{\rm in}}(\sigma^{x}-i\sigma^{y})/2$, we obtain
\begin{eqnarray}
    \Dot{\rho} &=& \gamma_{2}(L_{\rm in}\rho L^{\dagger}_{\rm in} - \frac{1}{2}\{ L^{\dagger}_{\rm in} L_{\rm in},\rho\})\\
    &=& \begin{bmatrix}
        -\gamma_{\rm in}p_{0} & -\gamma_{\rm in}c/2 \\
        -\gamma_{\rm in}c^{*}/2 & \gamma_{\rm in}p_{0}
    \end{bmatrix} ,
\end{eqnarray}
and the infinitesimal time evolution under $\mathcal{L}_{\mathrm{in}}$ as
\begin{eqnarray}
    \rho(dt) &=& e^{\mathcal{L}_{\mathrm{in}}dt} \\ &\approx & \rho(0) + dt \Dot{\rho} \\
    &=& \begin{bmatrix}
        p_{0}-dt \gamma_{\rm in}p_{0} & c-dt \gamma_{\rm in}c/2 \\
        c^{*}-dt \gamma_{\rm in}c^{*}/2 & p_{1}+dt\gamma_{\rm in} p_{0}\label{eq:Density-matrix-using-ancilla-Lin}
    \end{bmatrix}
\end{eqnarray}
Comparing Eq.~\eqref{eq:Density_matrix_after_applicaion_of_U_ancilla} and Eq.~\eqref{eq:Density-matrix-using-ancilla-Lin}
\begin{eqnarray}
    \sin^{2}\theta &=& dt \gamma_{\rm in}\\
    \theta &=& \arcsin (\sqrt{dt \gamma_{\rm in}})
\end{eqnarray}
Note that these quantum circuits require $dt \gamma_{\mathrm{in}} < 1$ and $dt \gamma_{\mathrm{out}} < 1$.

\begin{figure}
    \centering
    \includegraphics[width=0.9\linewidth]{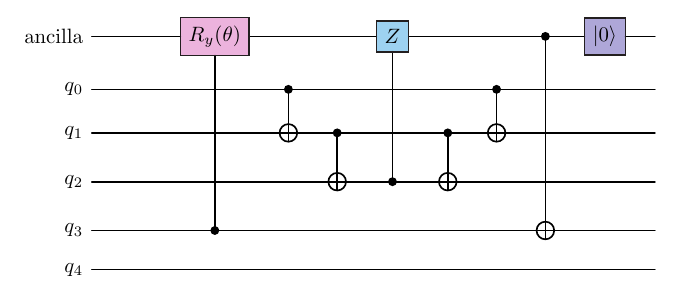}
\includegraphics[width = \linewidth]{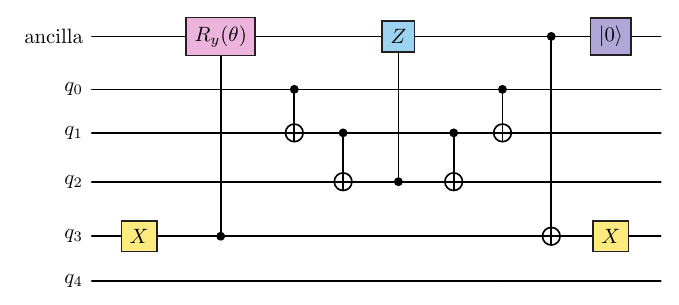}        \caption{Example quantum circuit to implement $c_{3}$ (top) and $c^{\dagger}_{3}$ (bottom) for a five qubit system $(L = 5)$. The ladder of CNOT gates represents the term $\sigma^{z}_{2} \otimes \sigma^{z}_{1} \otimes \sigma^{z}_{0}$. Starting from a computational basis state $|q_{0} q_{1} q_{2} q_{3} \rangle = |a, b, c, d\rangle$, the first CNOT produces $|a, b \oplus a, c, d\rangle$. The second CNOT produces $|a, b \oplus a, c \oplus b \oplus a, d\rangle$. Thus, the qubit $q_{2}$ stores the parity of the first three qubits $q_{2} = a \oplus b \oplus c$. After this a controlled-$Z$ gate is applied between the ancilla qubit and the qubit $q_{2}$. This gate introduces a phase $(-1)^{a_{\mathrm{anc}} (a \oplus b \oplus c)}$. In the ancilla construction described earlier for the $\sigma^{-}$ or $\sigma^{+}$ implementation, the jump occurs in the branch when the ancilla is in the state $|1\rangle$. Consequently, the controlled-$Z$ produces the phase $(-1)^{a + b + c}$, which is precisely the phase generated by the operator $Z_{0} Z_{1} Z_{2}$. After the phase is applied, the CNOT ladder is reversed, which restores the original values of the system qubits.}
    \label{fig:L_in_L_out_operator_circuit_diagrams}
\end{figure}

We now show the implementation of the jump operators when the Jordan-Wigner string $\Pi^{L}_{k < n} \sigma^{z}_{k}$ is present. The string part $\prod_{k < n} \sigma^{z}_{k}$ ensures that the fermionic anticommutation relations are satisfied. The action of this factor on a computational basis is simple
\begin{eqnarray}
    \prod_{k < n} \sigma^{z}_{k} |q_{0} q_{1} \cdots q_{n-1}\rangle &=& (-1)^{\sum_{k < n} q_{k}} |q_{0} q_{1} \cdots q_{n-1}\rangle \nonumber
\end{eqnarray}
where $q_{j} = 0$ or $1$.
We implement this parity computing string using a ladder of CNOT gates combined with a controlled $Z$ operation, which is illustrated in the figure
~\ref{fig:L_in_L_out_operator_circuit_diagrams} for a $L = 5$ qubit system.

\section{Derivation of the tunneling rates $\gamma_{\mathrm{in}}, \gamma_{\mathrm{out}}$ using Fermi's golden rule} \label{sec:gamma_in_gamma_out_from_Fermi's_golden_rule}
In this section, we derive the tunneling rates appearing in the jump operators. Similar first principle calculations have been performed in earlier works \cite{b8tq-169m, PhysRevB.94.155142} . We start by dividing the total Hamiltonian into system $+$ bath $+$ interaction terms
\begin{eqnarray}
    H &=& H_{S} + H_{B} + H_{I}
\end{eqnarray}
where
\begin{eqnarray}
    H_{S} &=& \epsilon d^{\dagger} d \\
    H_{B} &=& \sum_{k} \epsilon_{k} c^{\dagger}_{k} c_{k} \\
    H_{I} &=& \sum_{k} (J_{k} d^{\dagger} c_{k} + J^{*}_{k} c^{\dagger}_{k} d),
\end{eqnarray}
The fermionic operator $d^\dagger$ ($d$) creates (annihilates) an electron in a molecular orbital of the OPE-SMe molecule with energy $\epsilon_{d}$, and they satisfy $\{ d, d^\dagger \} = 1,d^2 = (d^\dagger)^2 = 0$.  The lead operators $c_k^\dagger$ ($c_k$) create (annihilate) an electron in a single-particle eigenstate $k$ with energy $\epsilon_{k}$ of the metallic lead, which is treated as a noninteracting fermionic reservoir, and they satisfy $\{ c_k, c_{k'}^\dagger \} = \delta_{k k'}$. The coupling $J_{k}$ describes tunneling between the molecular orbital and the lead state $k$. The interaction Hamiltonian contains two physical tunneling processes. The term $d^{\dagger} c_{k}$ annihilates an electron in the lead and creates it on the molecule. This corresponds to electron injection from the lead to the molecule. The term $c^{\dagger} d$ annihilates an electron on the molecule and creates it in the lead. This corresponds to electron extraction from the molecule into the lead. The leads are assumed to remain in thermal equilibrium at temperature $T$ and chemical potential $\mu$. Therefore, the lead occupation is described by the Fermi-Dirac distribution
\begin{eqnarray}
    f(\epsilon) &=& \frac{1}{e^{(\epsilon - \mu)/(k_{B}T)} + 1}
\end{eqnarray}
which gives the thermal averages
\begin{subequations}
\begin{eqnarray}
    \langle c_{k} c^{\dagger}_{k^{'}} \rangle &=& \delta_{k k^{'}} (1 - f(\epsilon_{k})) \\
    \langle c^{\dagger}_{k^{'}} c_{k} \rangle &=& \delta_{k k^{'}} f(\epsilon_{k}), 
\end{eqnarray}\label{eq:Thermal-average-formula}
\end{subequations}

We now derive the injection rate $\gamma_{\mathrm{in}}$. The bath operator associated with electron injection is
\begin{eqnarray}
    B &=& \sum_{k} J_{k} c_{k}
\end{eqnarray}
The correlation function for the reservoir is \cite{10.1093/acprof:oso/9780199213900.002.14006}
\begin{eqnarray}
    C_{\mathrm{in}}(t) &=& \langle B^{\dagger} (0) B(t)\rangle, \label{eq:Correlation_function_electron_from_environment_to_molecule}
\end{eqnarray}
To evaluate the time-evolved function $B(t)$, we use the Heisenberg equation of motion
\begin{eqnarray}
    \frac{d}{dt} c_{k}(t) &=& \frac{i}{\hbar} [H_{B}, c_{k}(t)] \\
    &=& \frac{i}{\hbar} \sum_{k^{'}} \epsilon_{k^{'}} [c^{\dagger}_{k^{'}}  c_{k^{'}}, c_{k}] \\
    &=& \frac{i}{\hbar} \sum_{k^{'}} \epsilon_{k^{'}} [c^{\dagger}_{k^{'}}  c_{k^{'}}, c_{k}] \\
    &=& \frac{i}{\hbar} \sum_{k^{'}} \epsilon_{k^{'}} (c^{\dagger}_{k^{'}} \{c_{k^{'}}, c_{k}\} - \{c^{\dagger}_{k^{'}}, c_{k}\}c_{k^{'}}) \\
    &=& 0 + \frac{i}{\hbar} (- \delta_{k k^{'}} c_{k^{'}}) \\
    &=& - \frac{i}{\hbar} \epsilon_{k} c_{k}(t) .
\end{eqnarray}
Solving this differential equation, we obtain
\begin{eqnarray}
    c_{k}(t) &=& e^{-i\epsilon_{k}t/\hbar} c_{k}(0) .
\end{eqnarray}
This gives
\begin{eqnarray}
    B(t) &=& \sum_{k} J_{k} e^{-i\epsilon_{k}t/\hbar} c_{k}
\end{eqnarray}
Substituting this in Eq.~\eqref{eq:Correlation_function_electron_from_environment_to_molecule}, we get
\begin{eqnarray}
    C_{\mathrm{in}}(t) &=&   \sum_{k^{'}} \sum_{k} J^{*}_{k^{'}} J_{k} \langle c^{\dagger}_{k^{'}}  e^{-i\epsilon_{k}t / \hbar}  c_{k} \rangle\label{eq:Correlation_function_after_time-evolution}
\end{eqnarray}
Using Eq.~\eqref{eq:Thermal-average-formula} in Eq.~\eqref{eq:Correlation_function_after_time-evolution}, we get
\begin{eqnarray}
    C_{\mathrm{in}}(t) &=& \sum_{k} |J_{k}|^{2} e^{-i \epsilon_{k}t / \hbar} f(\epsilon_{k}),
\end{eqnarray}
In the continuum limit, we can write this sum as an integral
\begin{eqnarray}
    \sum_{k} \rightarrow \int \rho_{\mathrm{lead}}(\epsilon) d\epsilon
\end{eqnarray}
where $\rho_{\mathrm{lead}}(\epsilon)$ is the density of states of the metallic lead. Thus 
\begin{eqnarray}
    C_{\mathrm{in}}(t) &=& \int e^{-i\epsilon t / \hbar} |J(\epsilon)|^{2} \rho_{\mathrm{lead}}(\epsilon) f(\epsilon) d\epsilon,
\end{eqnarray}
The Fourier transform of this correlation function is
\begin{subequations}
\begin{eqnarray}
    C_{\mathrm{in}}(\omega) &=& \int^{\infty}_{-\infty} e^{i \omega t} C_{\mathrm{in}}(t) dt \\
    &=& \int |J(\epsilon)|^{2} \rho_{\mathrm{lead}}(\epsilon) f(\epsilon) \int^{\infty}_{-\infty} e^{i(\omega - \epsilon/\hbar)t} dt \nonumber
\end{eqnarray}
\end{subequations}
Using
\begin{eqnarray}
    \int^{\infty}_{-\infty} e^{i(\omega - \epsilon/\hbar)t} dt &=& 2 \pi \delta(\omega - \epsilon/\hbar)
\end{eqnarray}
we obtain
\begin{eqnarray}
    C_{\mathrm{in}} &=& 2 \pi \hbar |J(\hbar\omega)|^{2}\rho_{\mathrm{lead}}(\hbar\omega)f(\hbar\omega)
\end{eqnarray}
This gives us the tunneling rate for injection as
\begin{eqnarray}
    \gamma_{\mathrm{in}} &=& \frac{1}{\hbar^{2}}C_{\mathrm{in}}(\omega)
\end{eqnarray}
Evaluating the rate at the molecular energy $\epsilon_{d} = \hbar \omega_{d}$, we get
\begin{eqnarray}
\gamma_{\mathrm{in}}(\epsilon_{d}) &=& \frac{2\pi}{\hbar}|J(\epsilon)|^{2}\rho_{\mathrm{lead}}(\epsilon_{d})f(\epsilon_{d})
\end{eqnarray}
This is the rate at which an electron tunnels from the lead into the molecular orbital.

We now derive the extraction rate $\gamma_{\mathrm{out}}$. In the extraction process, the molecule loses an electron and the lead gains one. The relevant process is described by the bath operator $B^{\dagger}$, and the corresponding correlation function is
\begin{eqnarray}
    C_{\mathrm{out}}(t) &=& \langle B(t) B^{\dagger} (0) \rangle
\end{eqnarray}
Using a similar procedure, we get
\begin{eqnarray}
    B(t) &=& \sum_{k} J_{k} e^{-i\epsilon_{k}t/\hbar} c_{k}
\end{eqnarray}
we obtain
\begin{eqnarray}
    C_{\mathrm{out}}(t) &=& \sum_{k^{'}} \sum_{k} J_{k}J^{*}_{k^{'}} e^{-i\epsilon_{k}t/\hbar}\langle c_{k}c^{\dagger}_{k^{'}}\rangle
\end{eqnarray}
Using the thermal average in Eq.~\eqref{eq:Thermal-average-formula}, we obtain
\begin{eqnarray}
    C_{\mathrm{out}}(t) &=& \sum_{k}|J_{k}|^{2}e^{-i\epsilon_{k}t/\hbar}[1-f(\epsilon_{k})]
\end{eqnarray}
Taking the continuum limit
\begin{eqnarray}
    C_{\mathrm{out}}(t) &=& \int |J(\epsilon)|^{2}\rho_{\mathrm{lead}(\epsilon})[1-f(\epsilon)]e^{-i\epsilon t/\hbar}d\epsilon
\end{eqnarray}
Using the Fourier transform, we obtain
\begin{eqnarray}
    C_{\mathrm{out}}(t) &=& 2\pi\hbar|J(\hbar\omega)|^{2}\rho_{\mathrm{lead}}(\hbar\omega)[1-f(\hbar\omega)]
\end{eqnarray}
The extraction rate is
\begin{eqnarray}
    \gamma_{\mathrm{out}}(\omega) &=& \frac{1}{\hbar^{2}} C_{\mathrm{out}}(\omega)
\end{eqnarray}
Evaluating this for $\epsilon = \epsilon_{d}$, we obtain
\begin{eqnarray}
    \gamma_{\mathrm{out}} &=& \frac{2\pi}{\hbar}|J(\epsilon_{d})|^{2}\rho_{\mathrm{lead}}(\epsilon_{d})[1-f(\epsilon_{d})]
\end{eqnarray}
The corresponding jump operators are
\begin{eqnarray}
    L_{\mathrm{in}} &=& \sqrt{\gamma_{\mathrm{in}}}d^{\dagger} \\
    L_{\mathrm{out}} &=& \sqrt{\gamma_{\mathrm{out}}} d
\end{eqnarray}
For transport through a molecule connected to two leads, the same derivation is applied separately to the left and right reservoirs. Thus, for lead $a=L, R$
\begin{subequations}
\begin{eqnarray}
\gamma^{a}_{\mathrm{in}}(\epsilon_{d}) &=& \frac{2\pi}{\hbar}|J(\epsilon)|^{2}\rho_{a}(\epsilon_{d})f(\epsilon_{d}) \\
\gamma^{a}_{\mathrm{out}} &=& \frac{2\pi}{\hbar}|J(\epsilon_{d})|^{2}\rho_{a}(\epsilon_{d})[1-f(\epsilon_{d})]
\end{eqnarray}
\end{subequations}
where
\begin{eqnarray}
    f_{a}(\epsilon_{n}) &=& \frac{1}{e^{(\epsilon_{n} - \mu_{a})/(k_{B}T)} + 1}
\end{eqnarray}
Here $\epsilon_{n}$ is the energy of the molecular orbital $n$, and $\rho_{a}(\epsilon_{n})$ is the density of available states in lead $a=L,R$ evaluated at that molecular orbital energy.
\section{Expressions for the currents $I_{\mathrm{measured}}$} \label{sec:Current-expectation-value}

In this section, we will derive the expressions for the \textit{in} and \textit{out} charge current in Eq.~\eqref{eq:I_measured}. We start with the representation of the number operator in the fermionic picture: $\hat{N} = \sum_{j} c^{\dagger}_{j} c_{j}$ where the sum is over the energy levels whose tunneling rates exceeds $\gamma_{\mathrm{thr}}$ The total current flowing through the system is given by
\begin{subequations}
    \begin{eqnarray}
        I_{\mathrm{total}} &=& e \frac{d \langle \hat{N} \rangle}{dt} = e \mathrm{Tr}[\frac{d \rho}{dt} \hat{N}]
    \end{eqnarray}
\end{subequations}
Using Eq.~\eqref{eq:Lindblad-equation-multilevel-tunneling-expression}, we obtain
\begin{eqnarray}
    && I_{\mathrm{total}} \\
    &=& e \sum_{n}\bigg(\gamma^{R}_{\mathrm{in}, n}\Tr[ (c^{\dagger}_{n} \rho _{n} - \frac{1}{2} c_{n} c^{\dagger}_{n} \rho - \frac{1}{2} \rho c_{n} c^{\dagger}_{n}) \sum_{j} c^{\dagger}_{j} c_{j}] \nonumber \\
    && + \;\gamma^{L}_{\mathrm{in}, n}\Tr[(c^{\dagger}_{n} \rho c_{n} - \frac{1}{2} c_{n} c^{\dagger}_{n} \rho - \frac{1}{2} \rho c_{n} c^{\dagger}_{n}) \sum_{j} c^{\dagger}_{j} c_{j}] \nonumber \\
    && + \gamma^{R}_{\mathrm{out}, m}\Tr[(c^{\dagger}_{m}  \rho c_{m} - \frac{1}{2} c^{\dagger}_{m} c_{m} \rho - \frac{1}{2} \rho c^{\dagger}_{m} c_{m}) \sum_{j} c^{\dagger}_{j} c_{j}] \nonumber \\
    && + \;\gamma^{L}_{\mathrm{in}, m}\Tr[(c_{m} \rho c^{\dagger}_{m} - \frac{1}{2} c^{\dagger}_{m} c_{m} \rho - \frac{1}{2} \rho c^{\dagger}_{m} c_{m}) \sum_{j} c^{\dagger}_{j} c_{j}]\Bigg) \nonumber\\
    &\equiv & I_{\mathrm{in}} + I_{\mathrm{out}}
\end{eqnarray}
Using the property of the Fermionic operators
\begin{eqnarray}
    c^{2} &=& (c^{\dagger})^{2} = 0, \{ c, c^{\dagger} \} = 1, c c^{\dagger} = 1 - N, \\ 
    N^{2} &=& N,
    N c = 0, c N  = c, c^{\dagger} N = 0,
\end{eqnarray}
and simplifying, we get
\begin{eqnarray}
    I_{\mathrm{in}} &=& e \sum_{n} \gamma^{L}_{\mathrm{in}, n} (1 - \langle \hat{N}_{n} \rangle) \nonumber \\
   && \hspace{1.2 cm} + e \sum_{m} \gamma^{R}_{\mathrm{in}, m} (1 - \langle \hat{N}_{m} \rangle)
\end{eqnarray}
Similarly for $I_{\mathrm{out}}$, we obtain
\begin{eqnarray}
   I_{\mathrm{out}}
   &=& - e \sum_{n} \gamma^{L}_{\mathrm{out}, n} \langle \hat{N}_{n}\rangle - e \sum_{m} \gamma^{R}_{\mathrm{out}, m} \langle \hat{N}_{m}\rangle
\end{eqnarray}
From these, we define our observables in Eq.~\eqref{eq:I_measured} and \eqref{eq:I_measured_equivalent}.
\section{Variational algorithm for the infinite-range dissipative Ising model}
In this section, we provide more details about the optimization procedure for the infinite-range dissipative Ising model. Due to the self-consistent nature of the Hamiltonian, the gradient calculation needs some extra caution, which we outline below.
\subsection{Gradient ascent for a general ansatz}
In this section, we discuss the calculation of the gradient for the two cost functions: $(1)$ Energy $(2)$ Frobenius norm. The parameters are updated according to the gradient ascent rule
\begin{eqnarray}
  \theta^{(n + 1)}_{k} &=& \theta^{(n)}_{k} + \eta \frac{\partial C}{\partial \theta_{k}}. \label{eq:gradient-ascent-formula}
\end{eqnarray}
Here $\eta > 0$ is the learning rate.

\textit{$(1)$ Energy cost function} ---
The energy cost function is $C[\rho_{f}] = \mathrm{Tr}[H_{f} \rho_{f}]$, where $H_{f} = x_{f} \sigma^{x} - \Delta \sigma^{z}$. We calculate the gradient with respect to the parameter $\theta$ as
\begin{subequations}
\begin{eqnarray}
    \frac{\partial C}{\partial \theta} &=& \Tr\left[\frac{\partial H_{f}}{\partial \theta} \rho_{f}\right] + \Tr\left[H_{f} \frac{\partial \rho_{f}}{\partial \theta}\right] \\
    &=& \Tr\left[\frac{\partial x_{f}}{\partial \theta} X \rho_{f}\right] + \Tr\left[H_{f} \frac{\partial \rho_{f}}{\partial \theta}\right] \\
    &=& \frac{\partial x_{f}}{\partial \theta} \Tr[ X \rho_{f}] + \Tr\left[H_{f} \frac{\partial \rho_{f}}{\partial \theta}\right] \\
    &=& \Tr\left[X \frac{\partial \rho_{f}}{\partial \theta}\right] x_{f} + \Tr\left[H_{f} \frac{\partial \rho_{f}}{\partial \theta}\right] \\
    &=& \mathrm{Tr}\left[(2 x_{f} \sigma^{x} - \Delta \sigma^{z}) \frac{\partial \rho_{f}}{\partial \theta}\right] .
\end{eqnarray}
\end{subequations}
We evaluate the derivative $\partial \rho_{f}/\partial \theta$ for $\theta \neq \theta_{x}$ using the chain rule:
\begin{eqnarray}
\frac{\partial e^{\theta \mathcal{L}}}{\partial \theta} &=& \mathcal{L} e^{\theta \mathcal{L}} . \label{eq:Derivative-for-theta-neq-theta_x}
\end{eqnarray}
For $\theta = \theta_{x}$, it requires some extra care because the quantity $x_{f}$ in layer $l$ depends on the previous layer $l-1$. Let $\rho_{L} = \mathcal{U}_{L} \circ \mathcal{U}_{L-1} \circ \cdots \circ \mathcal{U}_{l+1} \circ \mathcal{U}_{l} \circ \mathcal{U}_{l-1} \circ \cdots \circ U_{1} \rho_{0} =  \mathcal{U}_{L} \circ \mathcal{U}_{L-1} \circ \cdots \circ \mathcal{U}_{l+1} \circ \mathcal{U}_{l} \rho_{l-1}$. Using this, we evaluate the gradient as
\begin{subequations}
\begin{eqnarray}
    \frac{\partial \rho_{f}}{\partial \theta^{(l)}_{x}} &=& \frac{\partial }{\partial \theta^{(l)}_{x}} (\mathcal{U}_{L} \circ \mathcal{U}_{L-1} \circ \cdots \circ U_{l} \circ \rho_{l-1}) \nonumber\\
    &=& \mathcal{U}_{L} \circ \mathcal{U}_{L-1} \circ \cdots \circ \frac{\partial \mathcal{U}_{l}}{\partial \theta^{(l)}_{x}} \circ \rho_{l-1}\nonumber
    \\ 
    &=& \mathcal{U}_{L} \circ \mathcal{U}_{L-1} \circ \cdots \circ \mathcal{U}_{l+1} \circ \nonumber \\
    && e^{\theta^{(l)}_{z} \mathcal{L}_{z}}
    e^{\theta^{(l)}_{\mathrm{rel}} \mathcal{L}_{\mathrm{rel}}}
    e^{\theta^{(l)}_{y}\mathcal{L}_{y}}
    e^{\theta^{(l)}_{\mathrm{dep}} \mathcal{L}_{\mathrm{dep}}}\nonumber\\
    && \hspace{2.1 cm}
    \frac{\partial}{\partial \theta^{(l)}_{x}}e^{\theta^{(l)}_{x}\mathcal{L}_{x}} \rho_{l-1},
\end{eqnarray} \label{eq:Derivative-of-rho_f-with-respect-to-theta_x}
\end{subequations}
here $\rho_{l- 1}$ denotes the density matrix from the previous layer. To simplify the calculation, we write the $x-$rotation using unitary rotations:
\begin{eqnarray}
\frac{\partial}{\partial \theta^{(l)}_{x}} e^{\theta^{(l)}_{x}\mathcal{L}_{x}} \rho_{l-1} &=& \frac{\partial}{\partial \theta^{(l)}_{x}} U_{x}(\theta^{(l)}_{x}) \rho_{l-1} U^{\dagger}(\theta^{(l)}_{x}), \; \label{eq:Derivative_wrt_theta_x}
\end{eqnarray}
where $U_{x}(\theta^{(l)}_{x}) = e^{-i x^{(l)} \theta^{(l)}_{x} \sigma^{x}}$. Simplifying Eq.~\eqref{eq:Derivative_wrt_theta_x}, we obtain
\begin{eqnarray}
\frac{\partial}{\partial \theta^{(l)}_{x}} e^{\theta^{(l)}_{x}\mathcal{L}_{x}} \rho_{l-1} &=& -i x^{(l)} e^{-i x^{(l)} \theta^{(l)}_{x} \sigma^{x}}(\sigma^{x} \rho_{l-1} - \rho_{l-1} \sigma^{x}) \nonumber\\
&& \hspace{1 cm} e^{ix^{(l)} \theta^{(l)}_{x} \sigma^{x}}
\end{eqnarray}
Substituting in Eq.~\eqref{eq:Derivative-of-rho_f-with-respect-to-theta_x}, we get
\begin{eqnarray}
\frac{\partial \rho_{f}}{\partial \theta^{(l)}_{x}} &=& 
\mathcal{U}_{L} \circ \mathcal{U}_{L-1} \circ \cdots \circ \mathcal{U}_{l+1} \circ \nonumber\\
&& e^{\theta^{(l)}_{f} \mathcal{L}_{f}}
    e^{\theta^{(l)}_{\mathrm{rel}} \mathcal{L}_{\mathrm{rel}}}
    e^{\theta^{(l)}_{y}\mathcal{L}_{y}}
    e^{\theta^{(l)}_{\mathrm{dep}} \mathcal{L}_{\mathrm{dep}}} (-i x^{(l)} e^{-i x^{(l)} \theta^{(l)}_{x} \sigma^{x}}\nonumber\\
    && (\sigma^{x} \rho_{l-1} - \rho_{l-1} \sigma^{x}) e^{ix^{(l)} \theta^{(l)}_{x} \sigma^{x}}) \label{eq:Derivative-for-theta-equal-to-theta_x}
\end{eqnarray}

\textit{$(2)$ Frobenius norm cost function} ---
The Frobenius norm cost function is defined as
\begin{eqnarray}
    C[\rho] &=& \Tr[(\mathcal{L}\rho)^{\dagger} (\mathcal{L}\rho)]
\end{eqnarray}
Differentiating with respect to parameter $\theta$
\begin{subequations}
\begin{eqnarray}
\dv{C(\theta)}{\theta} &=& \mathrm{Tr}\bigg[ \frac{d}{d\theta} ((\mathcal{L}\rho)^{\dagger}) (\mathcal{L}\rho) + (\mathcal{L}\rho)^{\dagger}  \frac{d}{d \theta} (\mathcal{L}\rho)\bigg], \nonumber\\
&=& \mathrm{Tr}\bigg[ \bigg(\frac{d \mathcal{L}}{d \theta} \rho + \mathcal{L} \frac{d \rho}{d \theta} \bigg)^{\dagger} \mathcal{L} \rho \bigg] \nonumber\\
&& + \mathrm{Tr}\bigg[ (\mathcal{L} \rho)^{\dagger} \bigg( \frac{d \mathcal{L}}{d \theta} \rho + \mathcal{L} \frac{d \rho}{d \theta} \bigg) \bigg], \nonumber\\
&=& \mathrm{Tr}\bigg[ \bigg(\frac{d \mathcal{L}}{d \theta} \rho + \mathcal{L} \frac{d \rho}{d \theta} \bigg)^{\dagger} \mathcal{L} \rho \bigg], \nonumber\\
&=& \mathrm{Tr}\bigg[ \bigg(\bigg(\frac{d \mathcal{L}}{d \theta} \rho + \mathcal{L} \frac{d \rho}{d \theta} \bigg)^{\dagger} \mathcal{L} \rho \bigg)^{\dagger}\bigg], \nonumber\\
&=& 2 \mathrm{Re} \mathrm{Tr}\bigg[ \bigg(\frac{d \mathcal{L}}{d \theta} \rho + \mathcal{L} \frac{d \rho}{d \theta} \bigg)^{\dagger} \mathcal{L} \rho \bigg] .\nonumber
\end{eqnarray}
\end{subequations}
Using the chain rule, we obtain
\begin{eqnarray}
    \frac{d \mathcal{L}}{d \theta} \rho &=& \frac{d \mathcal{L}}{d x} \rho \frac{d x}{d \theta} = \frac{d \mathcal{L}}{d x} \rho \mathrm{Tr}[\sigma^{x} \frac{d \rho}{d \theta}] .
\end{eqnarray}
Using the definition of the Lindblad superoperator
\begin{eqnarray}
    \frac{d \mathcal{L}}{d x} \rho = -i [\sigma^{x}, \rho]
\end{eqnarray}
which gives us
\begin{eqnarray}
    \frac{d \mathcal{L}}{d \theta} \rho &=& -i \mathrm{Tr}[\sigma^{x} \frac{d \rho}{d \theta}][\sigma^{x}, \rho].
\end{eqnarray}
The derivative $\partial \rho/ \partial \theta$ is then calculated using Eq. 
 ~\eqref{eq:Derivative-for-theta-neq-theta_x} and Eq.~\eqref{eq:Derivative-for-theta-equal-to-theta_x}.

 \section{Detailed Calculation of Vacuum and Solvated OPE-SMe Models}
A gas-phase single-point energy calculation was performed on OPE-SMe at the B3LYP/6-31+G(d,p) level of theory~\cite{gaussian16}, yielding a system comprising 32 atoms, 142 electrons, and 420 basis functions. Distributed multipole analysis through rank~2 (monopole, dipole, and quadrupole) was carried out using GDMA~2.3.3~\cite{stone2005} on the converged SCF density from the formatted checkpoint file, with a switching radius of zero and a hydrogen radius parameter of 0.65~\AA. Bonded force-field parameters (bonds, angles, torsions, and out-of-plane bends) were derived from the Gaussian-optimized geometry using the TINKER valence utility~\cite{tinker}, while van der Waals parameters, atomic polarizabilities, and permanent multipoles were assigned from AMOEBA default values~\cite{ponder2010} for chemically equivalent atom types. The structure was subsequently minimized using L-BFGS optimization with an RMS gradient convergence criterion of 0.01~kcal/(mol$\cdot$\AA).

The solvated system was constructed using PACKMOL~\cite{packmol} by placing one OPE-SMe molecule at the center of mass of a cubic simulation box (edge length 42~\AA), surrounded by 500 THF molecules, with a minimum inter-molecular separation tolerance of 2.0~\AA, yielding a total system of 6,532 atoms. The THF force-field parameters were taken from the AMOEBA09 polarizable force field~\cite{ponder2010} and merged into a combined parameter file. The system was first minimized using the L-BFGS optimizer in TINKER-HP~\cite{tinker} with an RMS gradient convergence criterion of 1.0~kcal/(mol$\cdot$\AA). Molecular dynamics simulations were then performed in the NPT ensemble at 300~K using the Bussi velocity-rescaling thermostat~\cite{bussi2007} and a Monte Carlo barostat~\cite{chow1995}, with long-range electrostatics treated via Particle Mesh Ewald~\cite{essmann1995} with a real-space cutoff of 7.0~\AA\ and van der Waals interactions truncated at 7.0~\AA\ with a long-range dispersion correction. The equations of motion were integrated using the r-RESPA multiple time-step algorithm~\cite{tuckerman1992} with a 1~fs time step, and OPE-SMe solute atoms were held in place by a positional restraint of 100~kcal/(mol$\cdot$\AA$^{2}$) to maintain molecular planarity during solvent equilibration. A production trajectory of 100~ps with frames saved every 100~fs was collected for subsequent analysis.
 
Radial and spatial distribution functions (RDF and SDF) of the THF solvent around OPE-SMe were computed with TRAVIS~\cite{Brehm2011,Brehm2020} over the full production trajectory. The intermolecular RDF of THF oxygen (C$_4$H$_8$O) around the OPE-SMe solute (C$_{16}$H$_{14}$S$_2$) exhibits a broad first maximum at $r \approx 5.75$~\AA\ ($g \approx 1.29$), a shallow minimum at $r \approx 7.65$~\AA\ ($g \approx 0.76$, running coordination number $n \approx 0.025$), and a second diffuse peak at $r \approx 9.95$~\AA\ ($g \approx 1.31$). The modest peak heights, barely exceeding unity, confirm that the first solvation shell is weakly structured, consistent with the non-polar conjugated backbone offering no hydrogen-bond donor or acceptor sites. SDFs were computed in the molecular frame of OPE-SMe using C1, C7, and C11 as reference atoms, with an observation radius of 15~\AA\ and a bin resolution of 50~pm, revealing a diffuse and broadly distributed THF oxygen density without evidence of strong face-on stacking or tight directional coordination.

Despite this structural diffuseness, a collective orientation analysis reveals a statistically persistent orientational preference of the solvent. A THF molecule is counted as part of the solvation shell if its heavy-atom center of mass lies within $r_\mathrm{cut} = 8.0$~\AA\ of the nearest OPE-SMe backbone atom. For each shell molecule $i$, $\cos\theta_i$ is the cosine of the angle between the dipole-proxy vector $\vec{d}_i$ (ring-carbon COM $\to$ O, approximating the molecular dipole direction) and the radial vector $\vec{r}_i$ (THF COM $\to$ nearest backbone atom). Both vectors are evaluated under the minimum image convention, making the metric well-defined regardless of where a THF molecule sits along the extended backbone. The collective orientation score is then defined as

\begin{equation}
  S(t) = \frac{1}{N(t)}\sum_{i=1}^{N(t)}\cos\theta_i(t),
  \label{eq:score}
\end{equation}

\noindent where $S \to +1$ ($-1$) corresponds to all oxygens directed toward (away from) the backbone. Over 1000 trajectory frames, the distribution yields $\langle S\rangle = -0.219$ ($\sigma = 0.072$), with a mean shell occupancy of $\langle N\rangle = 34.3$ molecules (range 28--41). The score never exceeds $S = -0.005$, confirming that the system never achieves a genuinely O-toward collective state; the THF oxygen atoms persistently prefer to point away from the OPE-SMe $\pi$-scaffold. Two extreme snapshots: frame 929 ($S = -0.005$, near-isotropic, 17 of 33 shell THF molecules O-toward) and frame 92 ($S = -0.410$, strongly O-away, 5 of 29 shell THF molecules O-toward) were selected to maximize electrostatic contrast and used as starting geometries for the polarizable QM/MM single-point calculations described below.

Polarizable QM/MM single-point calculations were performed for two selected snapshots using LICHEM~\cite{lichem, gokcan2019lichem}, interfacing Gaussian~16~\cite{gaussian16} (QM) with TINKER~\cite{tinker} (MM, AMOEBA polarizable embedding). The QM region comprised all 32 OPE-SMe atoms treated at the B3LYP/6-31+G(d,p) level (neutral singlet; 71 occupied $\alpha$-spin orbitals; 402 spherical basis functions); the MM region comprised all 500 THF molecules (6,500 atoms). Long-range electrostatics were treated with the LREC scheme~\cite{fang2015,kratz2016lichem} (exponent 3) under periodic boundary conditions; LICHEM automatically reduced the requested 25~\AA\ cutoff to the minimum-image limit of half the box length, giving effective cutoffs of 20.66~\AA\ for the O-Away snapshot (frame~92, box $a = 41.32$~\AA) and 20.11~\AA\ for the O-Toward snapshot (frame~929, box $a = 40.22$~\AA). Of the 6,500 MM atoms, the O-Away configuration had 6,138 frozen and 362 active MM atoms ($\approx$28 THF molecules), while the O-Toward configuration had 6,118 frozen and 382 active MM atoms ($\approx$29 THF molecules). The AMOEBA permanent multipoles and induced dipoles of all MM atoms were converted to external point charges~\cite{fang2015,kratz2016lichem,gokcan2019lichem} and passed to Gaussian, resulting in an electrostatic embedding comprising 20,286 point charges for O-Away and 20,196 for O-Toward.

The molecule-electrode coupling strength is quantified by the sulfur-resolved conduction coupling~\cite{ratner2013,nitzan2003}

\begin{equation}
  J_S^{(k,k+1)} = \frac{|E_{k+1}-E_k|}{2}
                  \sqrt{w_S^{(k+1)}\,w_S^{(k)}},
  \label{eq:Js-calculation-definition}
\end{equation}

\noindent where $w_S^{(k)}=\sum_{\mu\in\mathrm{S\,3p}}|C_{\mu k}|^2$ is the sulfur $3p$ weight of MO~$k$, and $C_{\mu k}$ are the AO-MO coefficients extracted from the Gaussian formatted checkpoint files. The geometric-mean weight factor $\sqrt{w_S^{(k)}w_S^{(k+1)}}$ projects the eigenvalue splitting $|E_{k+1}-E_k|/2$ onto the sulfur-anchored conduction channel, isolating the molecule-lead contribution from the full molecular orbital spectrum.

For each solvated snapshot, the QM geometry and AMOEBA-derived MM point charges were parsed directly from the LICHEM-generated Gaussian input file and passed to PySCF~\cite{Sun2018pyscf,Sun2020pyscf}, which performed a fresh B3LYP/6-31+G(d,p) Kohn--Sham SCF calculation with polarizable MM embedding. An equivalent gas-phase calculation was also performed to isolate the pure electrostatic contribution of the solvent. A four-orbital active space comprising MOs 70, 71, 72, and 75 (0-based indices 69, 70, 71, 74; corresponding to HOMO$-$1, HOMO, LUMO, and LUMO$+$3) was selected based on physical relevance to the sulfur-anchored conduction channel. The effective one-body Hamiltonian was constructed by projecting the converged Fock matrix onto the active space and removing the double-counted Coulomb and exchange contributions from the occupied active orbitals. Two-electron repulsion integrals were transformed to the active MO basis and converted to physicists' notation. One- and two-body integrals over this space were used to construct the second-quantized fermionic Hamiltonian with IBM Qiskit-Nature ~\cite{qiskit_nature} for subsequent quantum transport analysis.
\clearpage
\onecolumngrid
\setcounter{section}{0}
\renewcommand{\thesection}{S\arabic{section}}
\renewcommand{\theequation}{S\arabic{equation}}
\renewcommand{\thefigure}{S\arabic{figure}}
\renewcommand{\thetable}{S\arabic{table}}
\setcounter{equation}{0}
\setcounter{figure}{0}
\setcounter{table}{0}

\title{Supplemental information}

\author{Sasanka Dowarah}
\email{sasanka.dowarah@utdallas.edu}
\affiliation{Department of Physics, The University of Texas at Dallas, Richardson, Texas 75080, USA}

\author{Abeda Sultana Shamma}
\email{Abeda.Shamma@UTDallas.edu}
\affiliation{Department of Physics, The University of Texas at Dallas, Richardson, Texas 75080, USA}
\affiliation{Department of Chemistry and Biochemistry, The University of Texas at Dallas, Richardson, Texas 75080, USA}

\author{Yazdan Maghsoud}
\email{ymaghsoud3@gatech.edu}
\thanks{Current address: School of Chemistry and Biochemistry, Georgia Institute of Technology, Atlanta, Georgia 30332, USA}
\affiliation{Department of Chemistry and Biochemistry, The University of Texas at Dallas, Richardson, Texas 75080, USA}

\author{G. Andrés Cisneros}
\email{andres@utdallas.edu}
\affiliation{Department of Physics, The University of Texas at Dallas, Richardson, Texas 75080, USA}
\affiliation{Department of Chemistry and Biochemistry, The University of Texas at Dallas, Richardson, Texas 75080, USA}

\author{Michael Kolodrubetz}
\email{mkolodru@utdallas.edu}
\affiliation{Department of Physics, The University of Texas at Dallas, Richardson, Texas 75080, USA}
\maketitle

\section{S1: Orbital energies, density of states, and tunneling matrix elements}
\begin{table}[H]
\centering
\caption{Orbital energies $E_k$ (eV) and sulfur $p$-projected weights $w_S^{(k)} = \text{S}_{15}\text{-}p + \text{S}_{16}\text{-}p$ for MOs 69-76 across all three environments, B3LYP/6-31+G(d,p). }
\label{tab:Ek_wS}
\setlength{\tabcolsep}{4pt}
\begin{tabular}{|c|cc|cc|cc|}
\toprule
 & \multicolumn{2}{c|}{Vacuum}
 & \multicolumn{2}{c|}{Away }
 & \multicolumn{2}{c|}{Toward} \\
MO
  & $E_k$ (eV) & $w_S^{(k)}$
  & $E_k$ (eV) & $w_S^{(k)}$
  & $E_k$ (eV) & $w_S^{(k)}$ \\
\midrule
69 & $-6.993$ & $0.463$ & $-9.089$ & $0.422$ & $-8.572$ & $0.423$ \\ \hline
70 & $-6.150$ & $0.660$ & $-8.109$ & $0.639$ & $-7.755$ & $0.653$ \\ \hline
71 & $-5.442$ & $0.345$ & $-7.456$ & $0.369$ & $-7.021$ & $0.351$ \\ \hline
72 & $-1.605$ & $0.067$ & $-3.674$ & $0.069$ & $-3.320$ & $0.062$ \\ \hline
73 & $-0.762$ & $0.023$ & $-2.857$ & $0.011$ & $-2.422$ & $0.012$ \\ \hline
74 & $-0.707$ & $0.021$ & $-2.803$ & $0.010$ & $-2.313$ & $0.000$ \\ \hline
75 & $-0.354$ & $0.229$ & $-2.422$ & $0.114$ & $-1.905$ & $0.125$ \\ \hline
76 & $-0.109$ & $0.696$ & $-1.932$ & $0.283$ & $-1.469$ & $0.328$ \\
\bottomrule
\end{tabular}
\end{table}

\begin{table}[H]
\centering
\caption{Sulfur-projected electronic couplings $J_S^{(k,k+1)}=\frac{|E_{k+1}-E_k|}{2}\sqrt{w_S^{(k)}\,w_S^{(k+1)}}$\,(meV) for adjacent MO pairs 69-76 across all three environments, B3LYP/6-31+G(d,p). $w_S^{(k)}$ uses $p$-orbital contributions only ($\text{S}_{15}\text{-}p + \text{S}_{16}\text{-}p$) from the Gaussian Atomic contributions section.}
\label{tab:Js}
\begin{tabular}{|c|c|c|c|}
\toprule
Pair
  & $J_S^{\mathrm{Vacuum}}$
  & $J_S^{\mathrm{Away}}$
  & $J_S^{\mathrm{Toward}}$ \\
  & (meV) & (meV) & (meV) \\
\midrule
$(69,70)$ & ${233.002}$ & ${254.450}$ & ${214.693}$ \\ \hline
$(70,71)$ & ${168.921}$ & ${158.543}$ & ${175.702}$ \\ \hline
$(71,72)$ & $291.681$ & $301.737$ & $261.058$ \\ \hline
$(72,73)$ & $16.546$ & $11.254$ & $12.247$ \\ \hline
$(73,74)$ & $0.604$ & $283.178$ & $0.000$ \\ \hline
$(74,75)$ & $12.240$ & $6.432$ & $0.000$ \\ \hline
$(75,76)$ & $48.906$ & $44.006$ & $44.142$ \\ \hline
\end{tabular}
\end{table}

\begin{table}[H]
\centering
\caption{HOMO-LUMO gap $\Delta E_{\mathrm{gap}} = E_{\mathrm{LUMO}} - E_{\mathrm{HOMO}}$
(MO\,72 $-$ MO\,71) across all three environments at the
B3LYP/6-31+G(d,p) level.}
\label{tab:homo_lumo_gap}
\begin{tabular}{|l|r|r|r|}
\toprule
Environment & $E_{\mathrm{HOMO}}$ (eV) & $E_{\mathrm{LUMO}}$ (eV) & $\Delta E_{\mathrm{gap}}$ (eV) \\
\midrule
Vacuum          & $-5.442$ & $-1.605$ & $3.837$ \\ \hline
O-Away     & $-7.456$ & $-3.674$ & $3.782$ \\ \hline
O-Toward   & $-7.021$ & $-3.320$ & $3.701$ \\
\bottomrule
\end{tabular}
\end{table}

\section{S2: Molecular orbital visualizations}
 
\begin{figure}[H]
    \centering
    \includegraphics[width=0.85\textwidth]{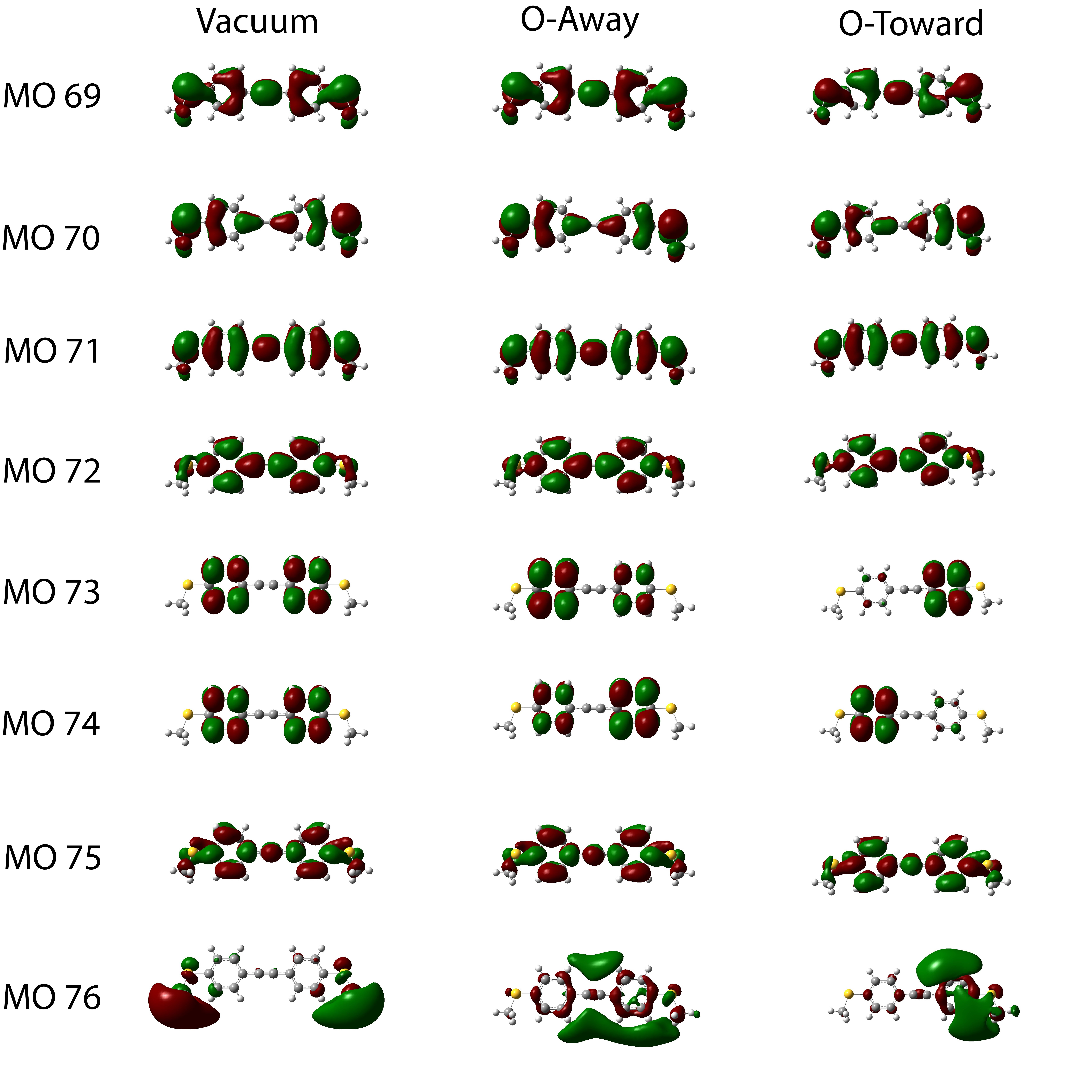}
    \caption{Isosurface plots of MOs 69-76 for OPE-SMe across three environments: Vacuum, O-Away, and O-Toward at B3LYP/6-31+G(d,p) level. Green and red lobes denote opposite phases of the wavefunction. Yellow atoms are sulfur; MOs 69-71 are fully occupied, MO 71 is the HOMO, and MO 72 is the LUMO.}
    \label{fig:MOs}
\end{figure}

\section{S3: Tunneling rates for OPE-SMe}
\begin{table}[H]
\centering
\caption{Tunneling rates for OPE-SMe in vacuum.}
\label{tab:tunneling_rates_vacuum}
\begin{tabular}{|l|c|c|c|c|}
\toprule
Orbital 
& $\gamma_{\mathrm{in}}^{L}\;(\mathrm{s}^{-1})$ 
& $\gamma_{\mathrm{out}}^{L}\;(\mathrm{s}^{-1})$ 
& $\gamma_{\mathrm{in}}^{R}\;(\mathrm{s}^{-1})$ 
& $\gamma_{\mathrm{out}}^{R}\;(\mathrm{s}^{-1})$ \\
\midrule
HOMO$-1$ & $3.42\times10^{14}$ & $0$ & $3.42\times10^{14}$ & $0$ \\ \hline
HOMO     & $9.40\times10^{13}$ & $0$ & $9.40\times10^{13}$ & $0$ \\ \hline
LUMO     & $0$ & $5.44\times10^{13}$ & $0$ & $5.44\times10^{13}$ \\ \hline
LUMO$+3$ & $0$ & $6.01\times10^{10}$ & $0$ & $6.01\times10^{10}$ \\ 
\bottomrule
\end{tabular}
\end{table}

\begin{table}[H]
\centering
\caption{Tunneling rates for OPE-SMe in the solvated-away configuration.}
\label{tab:tunneling_rates_solvated_away}

\begin{tabular}{|l|c|c|c|c|}
\toprule
Orbital 
& $\gamma_{\mathrm{in}}^{L}\;(\mathrm{s}^{-1})$ 
& $\gamma_{\mathrm{out}}^{L}\;(\mathrm{s}^{-1})$ 
& $\gamma_{\mathrm{in}}^{R}\;(\mathrm{s}^{-1})$ 
& $\gamma_{\mathrm{out}}^{R}\;(\mathrm{s}^{-1})$ \\
\midrule
HOMO$-1$ & $3.95\times10^{14}$ & $0$ & $3.95\times10^{14}$ & $0$ \\ \hline
HOMO     & $8.85\times10^{13}$ & $0$ & $8.85\times10^{13}$ & $0$ \\ \hline
LUMO     & $6.00\times10^{13}$ & $2.41\times10^{7}$ & $2.63\times10^{12}$ & $5.73\times10^{13}$ \\ \hline
LUMO$+3$ & $6.23\times10^{2}$  & $1.33\times10^{10}$ & $1.15\times10^{-5}$ & $1.33\times10^{10}$ \\ 
\bottomrule
\end{tabular}
\end{table}

\begin{table}[H]
\centering
\caption{Tunneling rates for OPE-SMe in the solvated-towards configuration.}
\label{tab:tunneling_rates_solvated_towards}
\begin{tabular}{|l|c|c|c|c|}
\toprule
Orbital 
& $\gamma_{\mathrm{in}}^{L}\;(\mathrm{s}^{-1})$ 
& $\gamma_{\mathrm{out}}^{L}\;(\mathrm{s}^{-1})$ 
& $\gamma_{\mathrm{in}}^{R}\;(\mathrm{s}^{-1})$ 
& $\gamma_{\mathrm{out}}^{R}\;(\mathrm{s}^{-1})$ \\ 
\midrule
HOMO$-1$ & $2.87\times10^{14}$ & $0$ & $2.87\times10^{14}$ & $0$ \\ \hline
HOMO     & $1.03\times10^{14}$ & $0$ & $1.03\times10^{14}$ & $0$ \\ \hline
LUMO     & $2.98\times10^{13}$ & $1.06\times10^{13}$ & $2.09\times10^{6}$ & $4.03\times10^{13}$ \\ \hline
LUMO$+3$ & $0$ & $1.72\times10^{10}$ & $0$ & $1.72\times10^{10}$ \\ 
\bottomrule
\end{tabular}
\end{table}
\bibliography{apssamp}
\end{document}